\documentclass[review]{elsarticle}

\usepackage{lineno,hyperref}
\modulolinenumbers[1]

\journal{Ocean Modelling}

\usepackage{url} 
\usepackage{soul}
\usepackage{graphicx}
\usepackage{latexsym}
\usepackage{amsmath}
\usepackage{amsfonts}
\usepackage{amssymb}
\usepackage{amsbsy}
\usepackage{amsgen}
\usepackage{color}
\usepackage{subeqnarray}
\usepackage{wasysym}

\usepackage{natbib}
\usepackage{rotating}
\usepackage{lineno}
\usepackage{hyperref}

\def\rmd{\mathrm{d}}

\def\be{\begin{equation}}
\def\ee{\end{equation}}

\begin{document}

\begin{frontmatter}
\title{Ice-thickness based scaling of wave attenuation in sea ice: Application and assessment of wave spectra}
\author{W. Erick Rogers$^1$}
\author{Jie Yu\corref{Corresponding author}$^1$}
\cortext[corresponding]{Corresponding author}
\ead{jie.yu4.civ@us.navy.mil}
\author{Jean Rabault$^2$, Ana Carrasco$^2$, and Malte M\"{u}ller$^{2,3}$}
\address{$^1$Naval Research Laboratory, Ocean Sciences Division, Stennis Space Center, MS 39529, USA\\
$^2$Norwegian Meteorological Institute, Oslo, Norway\\
$^3$University of Oslo, Oslo, Norway}

\begin{abstract}

This study discusses recent advances in modeling waves in sea ice in the U.S. Navy's regional modeling system. It is applied in the marginal seas of the eastern 
Arctic Ocean, including the Barents Sea, Kara Sea, parts of the Greenland Sea, Norwegian Sea, and waters north of Svalbard. 
The focus is to assess the skills of two formulations of wave attenuation by sea ice used operationally in WAVEWATCH III. 
Both are derived from large field datasets, one from the Arctic and the other from the Antarctic. The new model (IC4M9) describes wave attenuation 
depending on the ice thickness in association with the dependence on wave frequency, while the earlier default scheme (IC4M6) omits the dependence 
on ice thickness. The modeling results are evaluated against the satellite wave observations from SWIM/CFOSAT and the buoy measurements 
from the Svalbard Marginal Ice Zone 2024 Campaign (SvalMIZ-24). The comparisons with SWIM data validate the wave model skill in regions of open 
water or with light ice coverage. When evaluated against the SvalMIZ-24 data, the statistical performance of IC4M9 is substantially better than that of IC4M6, 
showing the influence of ice thickness on waves in the MIZ. Moreover, diagnosing systematic errors in the predictions by IC4M9, 
we find that the ice thickness field provided by the sea ice model CICE to the wave model is biased high in the MIZ, thus penalizing the performance of IC4M9
while not affecting the model IC4M6, which depends on frequency only. 

\end{abstract}
\end{frontmatter}


\section{Introduction}
\label{sec:intro}

The marginal ice zone (MIZ) is the region between open ocean waters and compact sea ice, where wave-ice interactions are the most dynamic. 
In the MIZ, the systems of  ocean, atmosphere, waves, and sea ice are dynamically coupled because of the strong interactions among them (e.g., Thomson et al., 2017; 
Boutin et al., 2022; Yang et al., 2024; M\"{u}ller et al., 2025), making forecasts especially challenging due to limitations in model representations of the complex physics. 
As the wave energy is  dissipated by sea ice, wave-induced ice breakup, ice drift, and the associated eddies and turbulent mixing, influence sea ice thermodynamics 
and dynamics in the MIZ, such as lateral melt and growth of ice, surface refreezing, floe welding, and floe size redistribution (e.g., Roach et al., 2018; 
Horvat et al., 2020;  Voermans et al., 2020; Boutin et al., 2021; Mokus and Montiel, 2022; Dumas-Lefebvre and Dumont, 2023).  
While the broken ice in the MIZ acts as a low-pass filter to effectively dissipate most wind wave energies, long waves of low frequencies 
continue penetrating deep into the pack ice, causing breakups 100s of km from the ice edge with low wave heights of a few tens of centimeters 
(Prinsenberg and Peterson, 2011; Kohout et al., 2016; Stopa et al., 2018; Nose et al., 2024). 

The Arctic has experienced accelerating loss of sea ice due to fast increasing temperatures in the region (Hayhoe et al., 2018). 
With increasingly open waters,  wave events in the Arctic MIZ are expected to be more frequent and extreme (Thomson and Rogers, 2014).  
The skills of forecasting systems in the Arctic can affect both long-term and short-term planning of human activities, including tactical decision-making by 
the U.S. Navy.  

In this paper, we are primarily concerned with the dissipation of wave energy by sea ice, which is a topic of research that 
has seen an intense activity in the recent years; see reviews in Squire et al. (1995), Squire (2018), Shen (2019), and Thomson (2022). 
This surge in interest is partly a response to the early implementations of parameterizations to represent the sea ice effects in operational wave 
modeling, such as the works of Doble and Bidlot (2013) for WAM at ECMWF (European Centre for Medium-Range Weather Forecasts), 
and Rogers and Orzech (2013) and Rogers and Zieger (2014) for WAVEWATCH III (Tolman, 1991; WW3DG, 2019). 
Prior to those works, sea ice was treated either very crudely or omitted entirely. Those efforts included various approaches to model the rate of wave 
attenuation by sea ice, primarily theoretical and empirical methods. 
The theoretical models involve practically non-measurable physical ice parameters, 
such as the effective viscosity and elasticity of broken ice in the MIZ (e.g., Wang and Shen, 2010; Mosig et al., 2015; Liu et al., 2020). 
Early evaluations of--and experiments with--these formulations in the U.S. Navy wave models argued against general adoption of these methods 
for operational use (Rogers et al., 2018a,b), as they were found to be difficult to calibrate and apply in operational modeling. 
Empirical formulas such as Doble et al. (2015) were also tested. The earliest such formulae, implemented in WAVEWATCH III by Collins and Rogers (2017), 
are derived from curve-fittings of data and often give large discrepancies when applied 
to different datasets without a recalibration, implying a lack of predictive skill across scales/conditions (Yu et al., 2022). 
The first U.S. Navy operational models opted for an empirical wave attenuation formula which has only the 
dependency on wave frequency and was derived from a field dataset in the Arctic MIZ (see Section \ref{sec:ic4m9}). 

It is of course very intuitive that ice conditions (i.e., ice concentration and thickness) have strong effects on the dissipation of wave energy. 
With respect to ice concentration, the common practice is to use it to weight the ice source term in the wave action equation in a spectral wave model. 
It is, however, not trivial to relate wave attenuation rate $k_i$ and ice thickness $h_{ice}$ via the data-based approaches. 
Since wave measurements are in the frequency domain, it is natural to estimate a $k_i$-$f$ relation (e.g. Meylan et al., 2014). Estimating the orthogonal 
dependence on $h_{ice}$ presents a special challenge, as it requires a sufficient amount of colocated measurements of waves with variety of ice thickness. 

In their attempt to understand the differences among existing theories on wave attenuation by ice, Yu et al. (2019) introduced an ice thickness based scaling 
and showed that when applied to field and lab datasets, scale collapse of data occurs, implying the relevance of $h_{ice}$. This was followed 
by Rogers et al. (2021b) and Yu et al. (2022) who analyzed a large field dataset using dimensionless variables and derived a new parameterization 
of wave attenuation, which upon returning to the dimensional form reveals a dependence on $h_{ice}$ in association with the dependence on wave 
frequency (see Section \ref{sec:ic4m9}). Rogers et al. (2025) documents the introduction of this new formula as the default option in the wave models 
used operationally by the U.S. Navy, WAVEWATCH III (WW3) and SWAN (Simulating Waves Nearshore, Booij et al., 1999).  
It has also been implemented in the ECMWF WAM model (Josh Kousal, personal communication; 
also see the presentation in \url{https://events.ecmwf.int/event/364/}), and in the Norwegian Meteorological Institute WAM model (Rabault et al., 2024).
In a coupled forecast system, $h_{ice}$ is  one of the ice fields routinely provided by numerical sea ice models to other model components. 
Establishing an accurate relationship between wave energy dissipation and ice thickness is important both for modeling waves and sea ice in the MIZ.  
In this paper, we present the first attempt to validate the application of this new formulation of wave attenuation in a U.S. Navy wave model WW3, 
evaluating the outcomes against observations.

For the description of earlier schemes, we refer the reader to WW3 User Manual (2019) and the references therein. The  new method of wave attenuation 
just mentioned above has recently been implemented into WW3 at NRL (US Naval Research Laboratory) and is publicly accessible 
(\url{https://github.com/NOAA-EMC/WW3}). Our focus here is to assess the skill of this new method in comparison with the earlier default 
scheme  in the Navy's  modeling system. Other plausible representations of $S_{ice}$ exist in the literature, such as the viscous layer, 
viscoelastic layer, viscoelastic beam, and porous viscoelastic models (Keller 1998, Wang and Shen 2010, Mosig et al. 2015, 
and Chen et al. 2019, respectively). These are not examined here; nor are models that estimate the dissipation caused by bottom friction at the 
ice-water interface (Stopa et al. 2016, Appendix B).  The scattering of wave energy by ice floes can also cause the attenuation of wave 
amplitudes (e.g., Dumont et al., 2011; Bennetts and Squire, 2012; Horvat and Tziperman, 2015; Meylan et al., 2021). 
This is the conservative effect of sea ice and not considered in this study. Fracturing of ice by waves is 
also not considered, but the reader is referred to the literature (e.g.,  Ardhuin et al., 2018; Williams et al., 2013; Boutin et al., 2018). 

The rest of the paper is organized as follows. Section \ref{sec:ic4m9} presents the parameterization of Yu et al. (2022). Section \ref{sec:ww3-barents} 
describes the modeling study which is the basis of our evaluation. Section \ref{sec:obs-swim} discusses comparisons of model results with satellite-derived 
wave spectra, primarily in areas of  approximately open water  in our model domain. In Section \ref{sec:obs-buoy}, we evaluate the predicted wave spectra 
against buoy measurements in and near the MIZ. In Section \ref{sec:hice-disscusion}, we discuss the accuracy of ice thickness field used in the 
prior sections. Summarizing remarks follow in  Section \ref{sec:summary}.

\section{Ice thickness-based scaling of wave attenuation in ice} 
\label{sec:ic4m9} 

For spectral wave models, e.g., WW3 and SWAN, the effect of sea ice is modeled by a source term 
$a_{ice}S_{ice}$ in the wave action equation, where $a_{ice}$ is the areal ice fraction, and $S_{ice}$ is the dissipation function, i.e., 
\be S_{ice} = -2 C_g k_i E. \label{eq:sice}\ee
Here, $C_g$ is the wave group velocity, $E$ is the spectral energy density, and $k_i$ is the spatial attenuation rate for the exponential decay of wave height. 
In WW3, several schemes for $k_i$ are available, including both theoretical models and data-based parameterizations (WW3DG, 2019). 
Rogers et al. (2021b) and Yu et al. (2022) developed a new formulation of $k_i$ that has a dependence on the thickness $h_{ice}$ of the broken-ice containing 
surface layer in association with the dependence on the wave frequency $f$; i.e., 
\be k_i(f,h_{ice}) = C h_{ice}^{1.25} f^{4.5}, \ \mbox{ where } \ C = 0.1274 (2\pi/\sqrt{g})^{4.5},  \label{eq:ic4m9}\ee  
and $g$ is the gravitational acceleration. For $g = 9.83 \mbox{ m/s}^2$ in polar regions, $C = 2.91$ (SI units). Formula (\ref{eq:ic4m9}) was obtained by 
fitting a large field dataset in the dimensionless plane $(\widehat\omega, \widehat{k}_i)$, where the normalized wave frequency and attenuation rate are defined as 
$\widehat\omega = (2\pi f) \sqrt{h_{ice}/g}$ and $\widehat k_i = k_i h_{ice}$, following the ice-thickness based scaling in Yu et al. (2019). 
In the plane $(\widehat\omega, \widehat{k}_i)$, a best fit $\widehat k_i = 0.1274\widehat\omega^{4.5}$ was found, which becomes formula (\ref{eq:ic4m9}) upon 
returning to the dimensional form. Thus, the dependencies on $f$ and $h_{ice}$ are related, with the power of $f$ being $n=4.5$ while the power of $h_{ice}$ 
being $n/2-1$.
The field dataset that was analyzed consists of 8957 attenuation rates estimated from the wave measurements from the ``Polynyas, Ice Production, 
and seasonal Evolution in the Ross Sea” (PIPERS) campaign in 2017 (Ackley et al., 2020) with the colocated $h_{ice}$ from the MIRAS (Microwave 
Imaging Radiometer using Aperture Synthesis)  radiometer onboard the European Space Agency’s SMOS (Soil Moisture and Ocean Salinity) satellite. 

As is pointed out by Yu et al. (2022), formula (\ref{eq:ic4m9}) represents a scaling similitude because of the normalization applied in its derivation. 
Indeed, without any re-calibration it can fit well other datasets from various independent field and laboratory studies (Yu et al., 2022, Figure~4). 
Being a curve-fitting, the physics behind (\ref{eq:ic4m9}) is not immediately apparent, but the recent theoretical study of wave boundary layer 
under sea ice gives a comparable solution, thus offering a  possible physical basis (Yu, 2022). 
Formula (\ref{eq:ic4m9}) is now available in the U.S. Navy wave models (WW3 and SWAN) and denoted within WW3 as the sea ice dissipation scheme IC4M9.  

Prior to the implementation of IC4M9, the default configuration in the Navy's operational modeling is to use the dissipation scheme IC4M6, 
in which $k_i$ depends only on $f$ and was derived by analyzing the extensive wave measurements 
from the Arctic Sea State field experiments funded by the Office of Naval Research. The IC4M6 parameterization $k_i(f)$ is described as a step 
function which is provided to WW3 using a Fortran namelist (Table \ref{table:ic4m6}). Details can be found in  Rogers et al. (2018a,b) on the formulation of 
the step function.
\begin{table} 
\caption{
$k_i(f)$ for IC4M6 provided to WW3 via the namelist, following Rogers et al. (2025). IC4FC are the maximum frequencies of the bins 
used in WW3, and IC4KI are the wave attenuation rates (1/m) in the corresponding bins. IC4FC = 99.0 flags $f > 0.40$ Hz.} 
\label{table:ic4m6} 
\begin{center}
\begin{tabular}{l l}
\hline
  IC4FC=\!&0.045, \, 0.055, \, 0.10,\, 0.15, \, 0.20, \, 0.25, \, 0.30, \, 0.40, \, 99.0\\
   IC4KI=\!&1.0e-06, \, 2.0e-06, \, 2.94e-06, \, 4.27e-06, \, 7.95e-06, \, 2.95e-05, \\
              &1.12e-04, \, 2.74e-04, \, 4.95e-04, \, 8.94e-04 \\
\hline
\end{tabular}
\end{center}
\end{table}

\begin{figure}
\centering\includegraphics[width=3.25in]{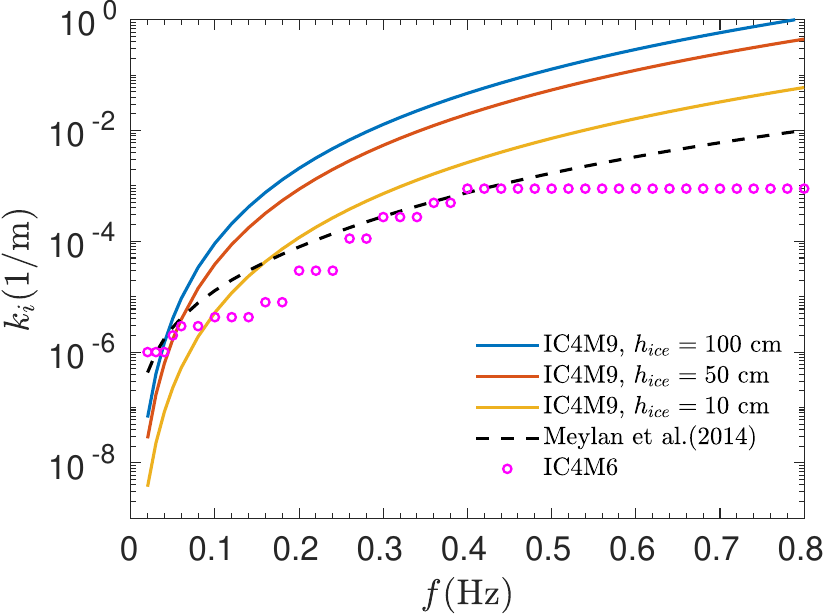}
\caption{Wave attenuation rate $k_i$ vs. $f$ given by different parameterizations.}
\label{fig:ki-f}
\end{figure}

The focus of this study is to assess the performances of IC4M9 and IC4M6 in high-latitude wave modeling within the Navy’s coupled regional modeling system,
COAMPS (Coupled Ocean Atmosphere Mesoscale Prediction System, Campbell et al., 2010; Smith et al., 2013; Doyle et al., 2014). 
Both schemes are available in COAMPS with IC4M9 being the new default, 
but they were derived from  observations under different sea ice conditions and in different polar oceans. 
The Arctic Sea State field dataset was from the wave measurements in frazil-pancake ice conditions in the autumn Beaufort Sea. 
The ice thickness is mostly around 5 to 10 cm based on the frazilometer data (Wadhams et al., 2018), while other estimate found the 
ice thickness less than 25 cm to be more credible (Rogers et al., 2016). The PIPERS dataset, however, was collected in the Antarctic winter, with 
the ice-fields being dominated by broken ice floes from new sheet ice (15 to 30 cm thick) and first-year ice (30 to 70 cm); see the summary in 
Rogers et al. (2021a, Table 1). Thus, IC4M6 is expected to be less dissipative than IC4M9 (Fig.~\ref{fig:ki-f}).
It is therefore interesting to see how well an empirical parametric model performs in a large-scale application beyond the conditions for which the model 
was originally derived.  Also included in Fig.~\ref{fig:ki-f} is the $f$-dependent only, binomial formula from Meylan et al. (2014), $k_i = a f^2 + b f^4$, which is  
based on the measurements from five on-ice wave sensors deployed in the Antarctic MIZ. 
The binomial formula is somewhat more dissipative than IC4M6, but considerably less dissipative than IC4M9 for moderate to high frequencies. 
For very low frequencies (e.g., $f < 0.06$ Hz), Meylan et al. (2014) and IC4M6 estimate the dissipation comparable to IC4M9 with $h_{ice} = 50 $ cm 
or thicker.  Recall that the colocated SMOS $h_{ice}$ for the PIPER dataset is capped at  50 cm because of the instrumentation saturation 
(Kaleschke et al., 2016).   

It should be mentioned that Montiel et al. (2025) evaluated 14 existing $S_{ice}$ schemes in WW3 (implemented by NRL, Ifremer, Clarkson University, 
and the University of Melbourne), including IC4M6 and Meylan et al. (2014) but not IC4M9, against the wave measurements from the PIPERS campaign. 
They concluded that most existing schemes (with default parameter values in WW3) provided too small 
wave energy damping relative to the measurements. In view of the large wave attenuation rates estimated  by IC4M9 (Fig.\,\ref{fig:ki-f}), we expect a strong 
performance of IC4M9 in the MIZ, especially at a heavy ice coverage (Section \ref{sec:obs-buoy}).

\section{WW3 modeling in the Barents Sea domain}
\label{sec:ww3-barents}
\begin{figure}
\centering\includegraphics[width=3.0in]{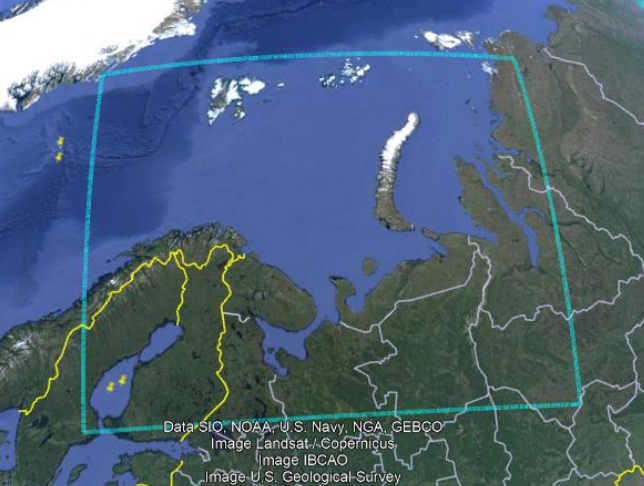}
\caption{The COAMPS/WW3 model domain for the extended Barents Sea region.}
\label{fig:model-domain}
\end{figure}

The model domain is shown in Fig.~\ref{fig:model-domain}. It includes the Barents Sea, Kara Sea, Svalbard, Franz Josef Land, the eastern 
Norwegian Sea, and the Gulf of Bothnia. Hereafter, we will refer it as the Barents Sea domain/grid. 
It is a 6-km Lambert Conformal grid, created using COAMPS utilities. Boundary forcing for COAMPS WW3 is derived from 
a global WW3 hindcast.  Surface winds are derived from the archives of the U.S. Navy’s global atmospheric model, NAVGEM (Navy Global 
Environmental Model, Hogan et al., 2014). 
Both $0.5^\circ$ and $0.18^\circ$ NAVGEM fields are used to  test the effect of wind resolution. For the uncoupled wave modeling in this study, 
the ocean surface current and water level fields are obtained from the database  of GOFS 3.1 (Global Ocean Forecast 
System), which is a $(1/12)^\circ$-resolution product. GOFS 3.1 is based on HYCOM (HYbrid Coordinate Ocean Model; Chassignet et al., 2003), 
two-way coupled to CICE v4.0 (Community Ice CodE; Hunke and Lipscomb, 2010)  and using the Navy Coupled Ocean Data Assimilation 
(NCODA). For the ice information input to WW3, the ice fraction field $a_{ice}$ can be taken from either 
the $0.18^\circ$ NAVGEM product or the $(1/12)^\circ$ GOFS 3.1 product, while the ice thickness field $h_{ice}$ (which is only used by IC4M9) is from GOFS 3.1. 
Both resolutions of $a_{ice}$ field are used to test the sensitivity of WW3 modeling to sea ice condition.  
A summary of the model configurations is in Table \ref{table:cc-4AB}. 
Two time periods are simulated, 10-20 February 2024 and 4-30 April 2024, giving respectively the durations of 7 and 23 days upon removing spin-up 
for comparisons with observations.  More details about model configuration can be found in Rogers et al. (2025).  
\begin{table}
\caption{Pearson correlation coefficients (CC) of model-observation comparisons. 
Runs A\# are for the time period of 10-20 February 2024, and runs B\# are for the time period of  4-30 April 2024. 
Column 2 lists the wind resolution from a 0.5$^\circ$ or 0.18$^\circ$ NAVGEM product. 
Column 3 lists the ice fraction resolution from a 0.18$^\circ$ NAVGEM product or the $(1/12)^\circ$ GOFS product. 
Column 4  indicates whether the ocean surface current is included. 
Column 5 shows the wave dissipation schemes that is used: IC4M9, IC4M6, or $S_{ice}=0$. The ice thickness used by IC4M9 is taken from GOFS. 
CC is computed using the colocations (630 for A\# and 1712  for B\#) between the model results and SWIM observations 
of $H_{m0}$ (total energy) and $H_{m0n}$ (energy in Band $n$).}\label{table:cc-4AB}
\begin{center}
\begin{tabular}{l l l c l | c c c c c}
\hline
          &                  &              &       &                     &\multicolumn{5}{c}{correlation coefficients}\\
    run &  wind res  & ice res  & ocn & {\small IC4} & {\small $H_{m01}$} & {\small $H_{m02}$} & {\small $H_{m03}$} & {\small $H_{m04}$} & {\small $H_{m0}$}\\
\hline   
A1 &{\small NAVGEM0.5}   &{\small GOFS1/12}      & y &{\small M9} & 0.64 &0.85 & 0.82 & 0.73 & 0.85\\
A2 &{\small NAVGEM0.18} &{\small GOFS1/12}      & y &{\small M9} & 0.70 & 0.91 & 0.87 & 0.75 & 0.90\\ 
A3 &{\small NAVGEM0.18} &{\small GOFS1/12}      & n &{\small M9} & 0.66 & 0.89 & 0.85 & 0.74 & 0.89\\
A4 &{\small NAVGEM0.18} &{\small GOFS1/12}      & n &{\small M6} & 0.67 & 0.90 & 0.86 & 0.83 & 0.91\\
A5 &{\small NAVGEM0.18} &{\small NAVGEM0.18} & n &{\small M6} & 0.68 & 0.89 & 0.86 & 0.84 & 0.91\\
A8 &{\small NAVGEM0.18} &{\small NAVGEM0.18} & y &0                 & 0.69 & 0.89 & 0.88 & 0.85 & 0.91\\
A9 &{\small NAVGEM0.18} &{\small NAVGEM0.18} & y &{\small M6} & 0.71 & 0.91 & 0.89 & 0.84 & 0.91\\
A11&{\small NAVGEM0.18}&{\small NAVGEM0.18}  & y &{\small M9}& 0.71 & 0.91 & 0.89 & 0.84 & 0.92\\[5pt]
B2 &{\small NAVGEM0.18} &{\small NAVGEM0.18} & y &{\small M9} & 0.77 & 0.96 & 0.97 & 0.94 & 0.98\\
B3 &{\small NAVGEM0.18} &{\small NAVGEM0.18} & y &{\small M6} & 0.78 & 0.97 & 0.97 & 0.94 & 0.98\\
B4&{\small NAVGEM0.18}&{\small NAVGEM0.18}   & y &0                 & 0.78 & 0.96 & 0.96 & 0.94 & 0.98\\
\hline
\end{tabular}
\end{center}
\end{table}

\begin{figure}
\centering\includegraphics[width=5.0in]{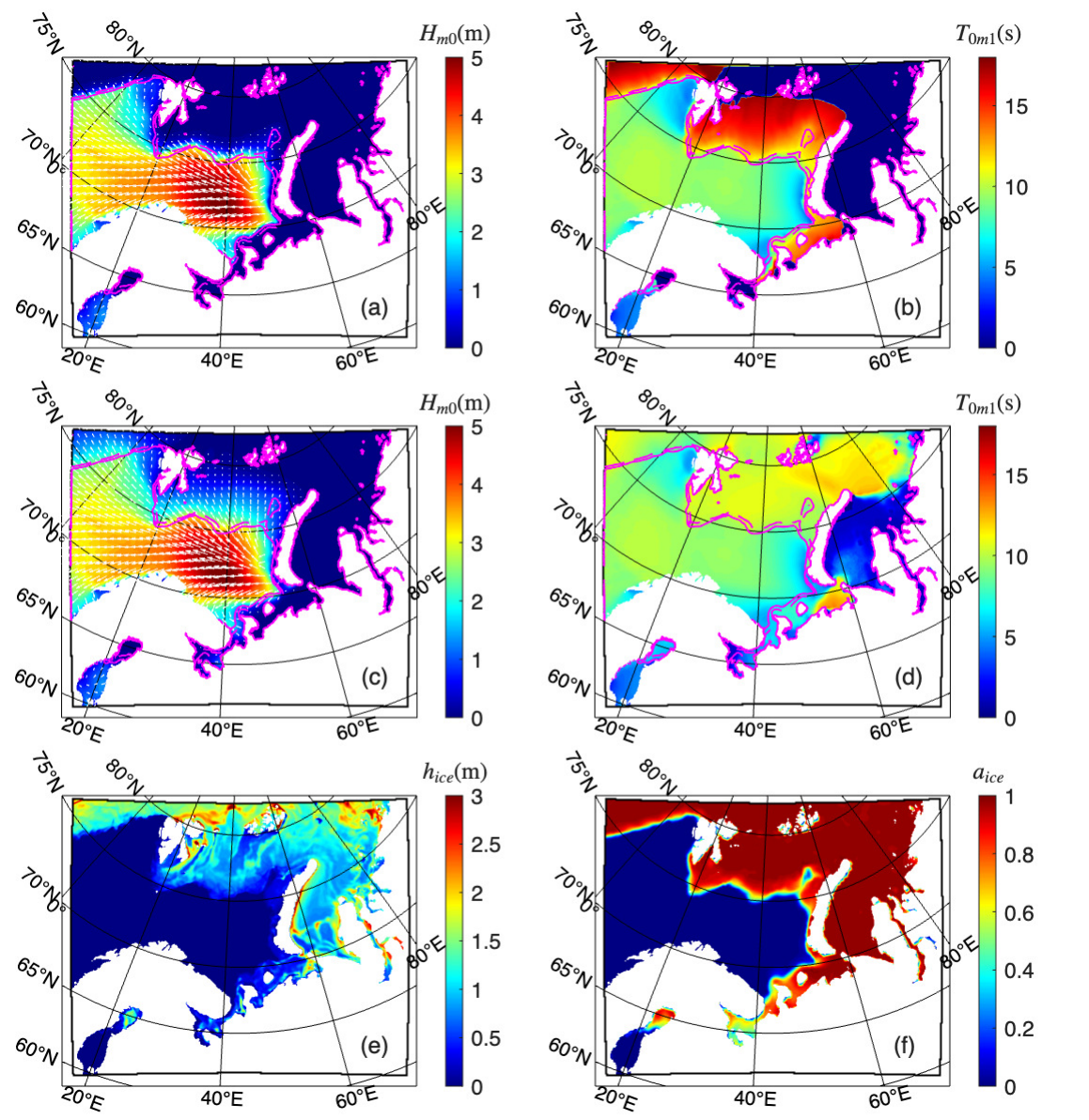}
\caption{(a) Significant wave height $H_{m0}$ superimposed by the peak wave direction (white arrows) and (b) mean wave period $T_{0m1}$ 
given by WW3 using IC4M9. (c) $H_{m0}$ and (d) $T_{0m1}$ given by WW3 using IC4M6. The magenta contours in (a)--(d) show the CICE ice 
concentration field of $a_{ice}=0.2$ (solid) and $a_{ice}=0.6$ (dashed). The sea ice fields input to WW3: (e) CICE ice thickness field $h_{ice}$ 
used only by IC4M9; (f)  CICE ice concentration field $a_{ice}$ used by both IC4M9 and IC4M6. The white areas are land. 
The valid time is 12-Apr-2024 0000 UTC. For the wind, ocean, and ice forcing, see runs B2 and B3 in Table \ref{table:cc-4AB}. }
\label{fig:ww3-wave-fields}
\end{figure}

Sample plots of the modeled wave fields are shown in Fig.~\ref{fig:ww3-wave-fields}. As expected, the wave energy rapidly 
diminishes once the waves propagate into the MIZ. Farther away from the ice edge, the energy spectrum increasingly down-shifts to low frequencies as the ice 
more effectively dissipates high frequency waves, acting as a low-pass filter. 
There are substantial differences in the wave fields between the model results using the new scheme IC4M9 and the previous default IC4M6 
when the ice concentration is sufficiently high; compare Figs.~\ref{fig:ww3-wave-fields}(a)--\ref{fig:ww3-wave-fields}(b) with 
Figs.~\ref{fig:ww3-wave-fields}(c)--\ref{fig:ww3-wave-fields}(d).  
Note that the significant wave height  $H_{m0}$ and mean  wave period $T_{0m1}$ are both masked to be zero when $H_{m0} < 0.001$ m. 
With IC4M9 the predicted wave energy becomes negligibly low at high ice concentrations, but with IC4M6 $H_{m0}$  can be higher than 2.0 m 
for $a_{ice} > 60$\%. Greater differences are seen in the mean wave period, with IC4M6 underestimating $T_{0m1}$ by about 5 s or more compared to 
that by IC4M9. This is due to the inability of IC4M6 to sufficiently dissipate wave energy at high frequencies (see Fig.~\ref{fig:ki-f}).  
It is also consistent with the findings in Montiel et al. (2025) that with the previously implemented schemes of wave attenuation, including IC4M6, 
over-estimations of significant wave height by WW3 in sea ice are accompanied by under-estimations of mean wave period.

\section{Evaluation against satellite wave observations}
\label{sec:obs-swim}

The SWIM (Surface Waves Investigation and Monitoring) instrument on the CFOSAT (China-France Oceanography SATellite) satellite provides wave spectra. 
This is used  to evaluate the model performance based on the total energy and energies within frequency bands. 
The geographic resolution of SWIM data is relatively coarse, with one wave spectrum corresponding to a 70 km x 90 km cell. 
The product is delivered in NRT (Near Real Time) within three hours of collection. It provides directional spectra, but in this study we only use the 
non-directional wave spectra. SWIM uses the radar cross-section of long waves to compute a spectrum of the modulation 
of the radar signal, which is in turn used to compute a slope spectrum $F(k)$, where $k$ is the wavenumber. We convert it to the frequency spectrum $E(f)$, 
assuming the deep-water dispersion relation. The range of  $k$ provided with the NRT Level 2 $F(k)$ spectra corresponds to  $f$ from 0.056  to 0.263 Hz. 
Of SWIM's six rotating beams, wave spectral information is available from beams with incidence 6$^\circ$, 8$^\circ$, 10$^\circ$. 
In this study,  we use the “combined” product. This is a weighted average of the wave slope spectrum of each incident beam, thereby reducing 
statistical noise (SWIM Products Users Guide, 2019). 
The 10$^\circ$ product is known to  have less ``speckle noise" than the other two beams, implying that it may be superior to the combined product 
(see, e.g., Hauser et al., 2021). We have compared the two products (i.e., 10$^\circ$ vs. combined) and found that the impact  on wave model validation 
is very slight. More details on SWIM and CFOSAT can be found in the literature (Hauser et al., 2021;  Le Merle et al., 2021; Aouf et al., 2019, 2021). 

The CFOSAT tracks provide almost global coverage within 13 days. There is a CFOSAT track every 90 minutes or so in the 
region of our interest, but not every track goes over the Barents sea model domain. Over the time, a large amount of data points can be located 
within the model domain. For each SWIM data point identified, we first interpolate WW3 2D spectra to the time and position of that point, 
then calculate the directional integration to obtain the co-located WW3 1D spectrum. Statistics of the model-observation comparison are computed 
using those colocations.

\begin{table} 
\caption{Frequency bands used to analyze the model-observation comparisons in Sections \ref{sec:obs-swim} and \ref{sec:obs-buoy}. 
A band is defined by $f_1<f<f_2$. }\label{table:f-bands}
\begin{center}
\begin{tabular}{l c c c c }
\hline
    & Band 1 &  Band 2 & Band 3 & Band 4 \\
\hline   
\multicolumn{5}{l}{\em SWIM/CFOSAT data:} \\   
$f_1$(Hz) & 0.056   & 0.080 & 0.110 & 0.150 \\
$f_2$(Hz) & 0.080  & 0.110  & 0.150 & 0.263 \\[5pt] 
\multicolumn{5}{l}{\em SvalMIZ-24 buoy data:}\\
$f_1$(Hz) & 0.044   & 0.070 & 0.110 & 0.150 \\
$f_2$(Hz) & 0.070  & 0.110  & 0.150 & 0.307 \\
\hline
\end{tabular}
\end{center}
\end{table}

The total energy is characterized by the significant wave height 
\be  H_{m0} = 4 \sqrt{\int_{f_{min}}^{f_{max}}\! E(f) \,\rmd f},  \label{eq:Hm0}\ee
where $f_{min}=0.056$ Hz and $f_{max}=0.263$ Hz for the SWIM data. We divide this frequency range into four bands 
(Table \ref{table:f-bands}) and quantify the energy within a band $f_1< f< f_2$ using the wave height, 
\be H_{m0n} = 4 \sqrt{\int_{f_{1}}^{f_{2}}\! E(f) \,\rmd f},  \label{eq:Hm0n}\ee 
where the subscript $n$ denotes the band number. The frequency range in WW3 modeling is much wider, ranging from 0.038 to 0.73 Hz. 
For the purpose of comparison, we truncate the modeled spectra to match the frequency limits of the observations.

The primary objective of the evaluation against SWIM observations is to assess the impacts on modeling results of various factors, 
including  the effects of ocean surface currents, resolutions of the wind and ice fraction forcing, and ice dissipation parameterization $S_{ice}$.  
A series of numerical experiments were carried out for the short time period of 10-20 February 2024. 
For the 7-day duration (upon removing spin-up), 630 colocations can be identified between the modeling results and SWIM/CFOSAT observations. 
These are used to compute the Pearson correlation coefficient (CC) of the model-observation comparison for each model run.  
The correlation statistics are  compared in Table \ref{table:cc-4AB}. A few points are noted: 
\begin{itemize}
\item[(a)] The higher resolution wind forcing has the largest positive impact on the model skill. When the  wind resolution is increased from 0.5$^\circ$ 
to 0.18$^\circ$, CC is improved in all frequency bands; compare run A1 with  A2 in Table \ref{table:cc-4AB}. 
\item[(b)] Including ocean surface currents also noticeably improves the model skill, increasing CC in all bands though the impact is not as strong as that 
caused by the wind resolution; compare run A3  with A2  in Table \ref{table:cc-4AB}. 
\item[(c)] For this test period, neither the resolution of ice fraction forcing (compare runs A4 and A5) nor inclusion/omission of ice thickness 
(compare runs A9 and A11) shows any statistical significance impacting the modeling results. In particular, when the ice dissipation function is turned 
off (i.e., $S_{ice}=0$), CC is only slightly decreased for Bands 1, 2, and 3 (compare run A8 with runs A9 and A11). 
\end{itemize}

\begin{figure}
\centering\includegraphics[width=4.75in]{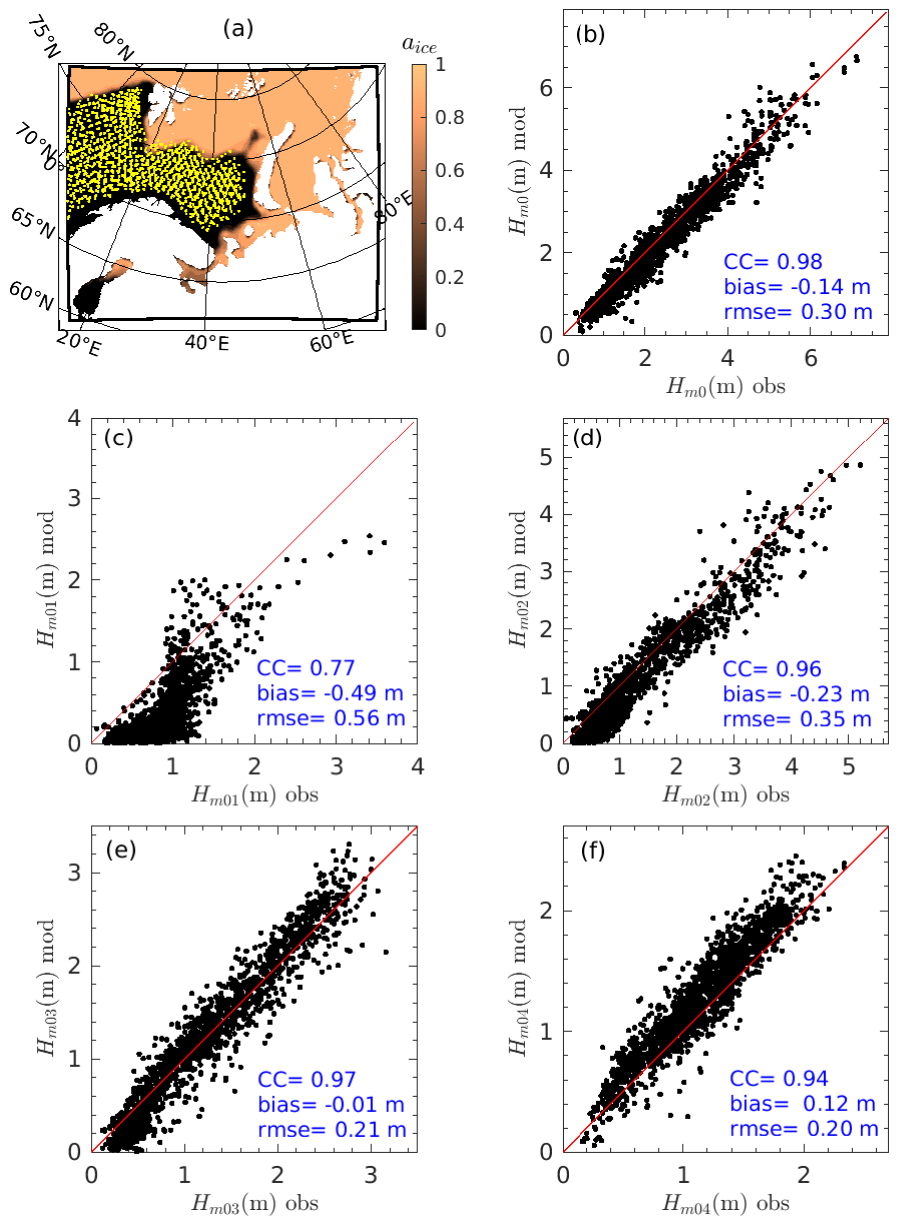}
\caption{(a) Locations of 1712 data points (yellow dots) colocated between the SWIM/CFOSAT observations and WW3 model fields during the time period 
of 7 to 30 April 2024. The background color scales show the CICE $a_{ice}$ field at 12-Apr-2024 0000 UTC. 
Scatter plots comparing the WW3 results vs. SWIM data  for the period of 7 to 30 April 2024: 
(b) $H_{m0}$ for the total energy; (c)--(f) $H_{m0n}$ for the energy Band $n=1,2,3,4$. The bands are listed in Table \ref{table:f-bands}. 
The statistics are computed for the 1712 colocations. The model results are from run B2 in Table \ref{table:cc-4AB}. }
\label{fig:scatter-plots-4B}
\end{figure}

While (a) and (b) are expected, the outcomes in (c) indicate that the wave modeling results are less sensitive to the effects of sea ice at those 
colocations for the test period of 14-20 February 2024.  Using the much larger number of 1712 colocations from the test period of 7-30 April 2024, 
we again find that the statistical correlations are insensitive to the choices of $S_{ice}$ (see runs B2, B3 and B4 in Table \ref{table:cc-4AB}).  
This is due to the fact that the SWIM observations are largely in regions of open water, with only a small portion of the data in areas of relatively high 
sea ice concentration; see Fig.~\ref{fig:scatter-plots-4B}(a). 
The SWIM L2 product flags spectra in ice as invalid, based on ice analyses 
from ECMWF.  While it was less ideal for evaluating the skill of modeling waves in sea ice than we had hoped,  the above comparison to SWIM 
observations did offer useful information on the accuracy of WW3 modeling in approximately open water in the Barents Sea domain:  
The CC values are high (near or above 0.90) for $f> 0.08$ Hz and somewhat reduced for $f < 0.08$ Hz. 

The scatter plots are shown in Figs.~\ref{fig:scatter-plots-4B}(b)--\ref{fig:scatter-plots-4B}(f) for the 1721 colocations during 7-30 April 2024. 
The model results from the run using IC4M9 are shown. 
At low frequencies, the modeling results tend to underestimate the observations, with negative biases of $-0.49$ m 
and $-0.23$ m in the modeled $H_{m01}$ and $H_{m02}$, respectively. For the high frequencies in Band 4, the modeled wave energies are biased slightly high 
against the SWIM observations. This is presumably due to the accuracy of WW3 model physics in open water, since the model results are insensitive to 
the effects of sea ice (Table \ref{table:cc-4AB}).  
As discussed in Hauser et al. (2021), parasitic peaks at low frequencies can occur in the SWIM spectra, which are likely due to  the noise floor 
when converting slope spectra to height spectra, and may not be well filtered. This can also contribute to the negative biases in the modeled $H_{m01}$ 
and $H_{m02}$ just mentioned. 

\section{Evaluation against  buoy wave observations}
\label{sec:obs-buoy}

To evaluate the performance of ice dissipation parameterization $S_{ice}$, we compare the model results with a dataset of buoy wave measurements in the 
Arctic MIZ.  In April 2024, the Norwegian Meteorological Institute (MET Norway) conducted the Svalbard Marginal Ice Zone 2024 Campaign (SvalMIZ-24) 
in the north of the Svalbard Archipelago. During the main observation period, there were cold-air outbreaks and subsequent warm-air intrusion, accompanied by 
strong winds and waves which caused a major transformation of the MIZ in just a few days (M\"{u}ller et al., 2025). 
The OpenMetBuoys (Rabault et al., 2022) were deployed directly on the ice. 
They are of an inexpensive, open-source design and capable of continuing measurements after falling into water, though they are not built to withstand 
strong crushing forces such as might happen when caught in interstitial water between large floes. 
Each buoy recorded non-directional wave spectra,  as well as temperatures of the air, snow, and ice from four thermistors (M\"{u}llet et al., 2024). 
A list of the contents of the data file can be found in Rogers et al. (2025).  The buoys have identification numbers KVS-01 to  KVS-35, 
and the measurements are available from 33 of the 35 buoys. Herein, we primarily refer to the buoys by their position-numbers in the 
data file, 1 to  33, rather than their IDs. 

\begin{figure}
\centering
\includegraphics[width=2.5in]{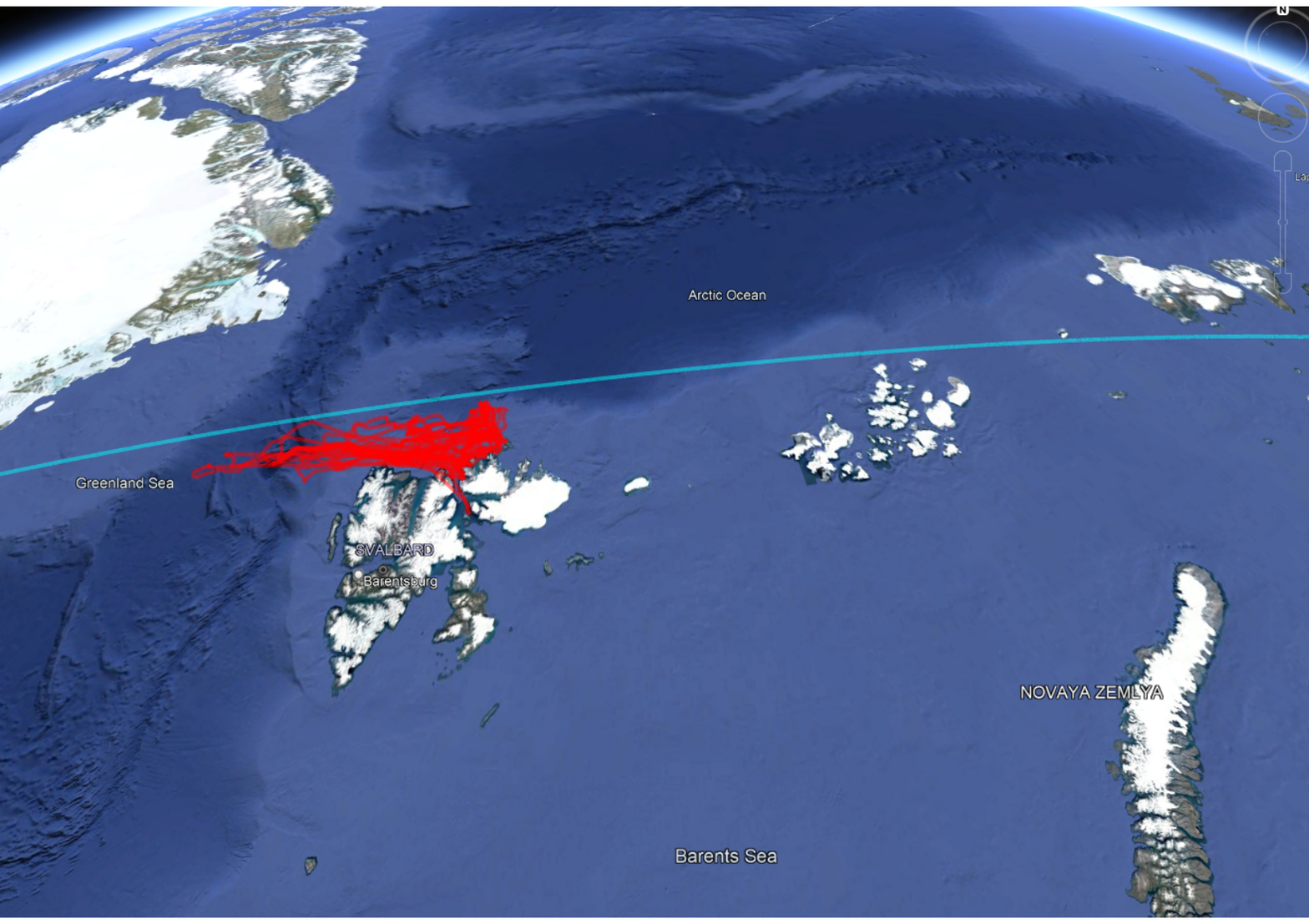}\quad
\includegraphics[width=2.0in]{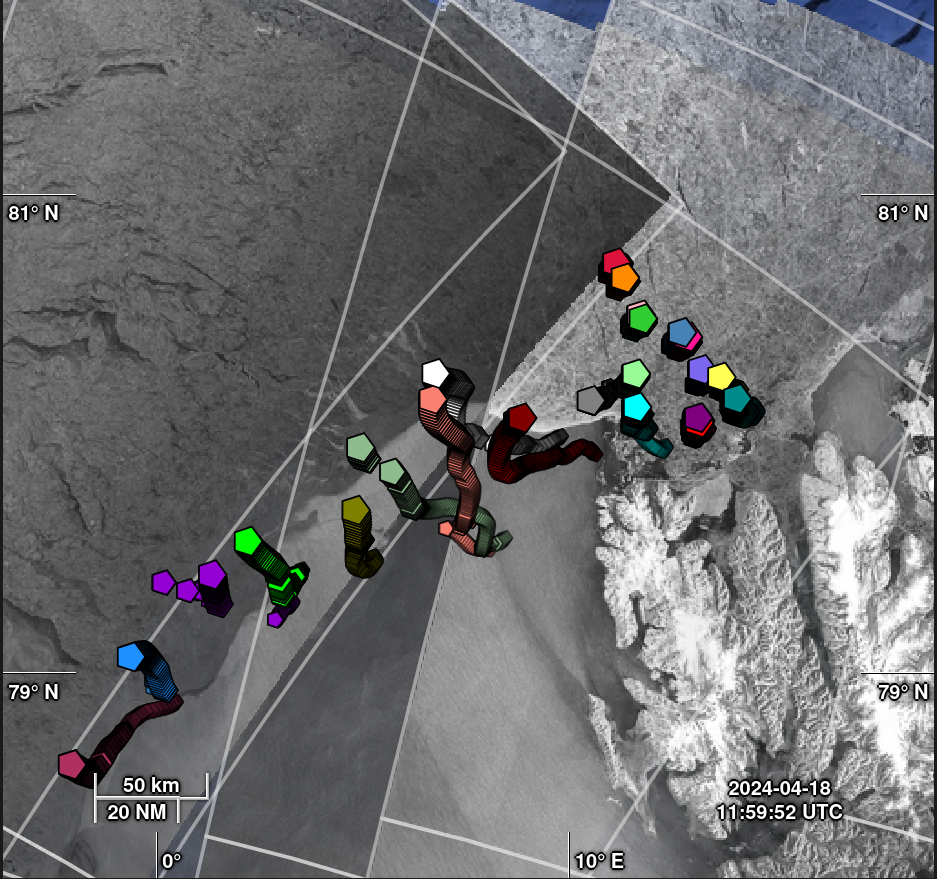}
\caption{(Left) Buoy tracks during the Svalbard Marginal Ice Zone 2024 experiment, 7 to 30 April 2024. The blue line is the northern boundary of the 
WW3 Barents Sea grid. (Right) Illustration of the SvalMIZ-24 deployment, visualized through the OVL portal. It shows the positions of the OMBs 
during the early phases of deployment. The background shows a snapshot of Synthetic Aperture Radar (SAR) images. 
The data can be browsed interactively at: \url{https://odl.bzh/1-TcoWKp}.} 
\label{fig:buoy-tracks}
\end{figure}

There are a total of 4273 buoy-derived wave spectra, $E(f)$, with $f_{min} = 0.0439$ Hz and 
$f_{max} = 0.307$ Hz. This is a broader frequency range (at either end) than the SWIM spectra (Section \ref{sec:obs-swim}), but still more restricted 
than the wave model. The data start at 0030 UTC 07 April 2024 and end at 2230 UTC 30 April 2024. Each buoy recorded its own time information and 
the corresponding position data (latitude, longitude), thus informing the time and position of each wave measurement. 
Waves were measured every two hours. 
The buoys were deployed predominately within the MIZ in the north and west of Svalbard, though many drifted into open water during the 
experiment (Fig.~\ref{fig:buoy-tracks}). The four frequency bands used to analyze the buoy data  are listed in Table \ref{table:f-bands}. 
The buoys trajectories, alongside other layers such as Synthetic Aperture Radar (SAR) imaging, can be browsed interactively from the Ocean 
Virtual Laboratory (OVL; see Collard et al., 2015) portal, and a snapshot of buoy trajectories and SAR background is visible in Fig.~\ref{fig:buoy-tracks}.

\begin{sidewaystable}
\caption{Biases (model $-$ observations) in $H_{m0}$ and $H_{m0n}$. In each frequency range, data are sorted 
from high to low observed wave energy (Obs $H$) and classified based on the relative  biases:  
Category I (light blue), $|\mbox{bias}|_{\mbox{\footnotesize IC4M6}} < |\mbox{bias}|_{\mbox{\footnotesize IC4M9}}$; 
Category II (yellow), $|\mbox{bias}|_{\mbox{\footnotesize IC4M9}} < |\mbox{bias}|_{\mbox{\footnotesize IC4M6}}$. 
The four frequency bands are defined in Table \ref{table:f-bands}. }\label{table:bias}
\centering
\includegraphics[width=7.0in]{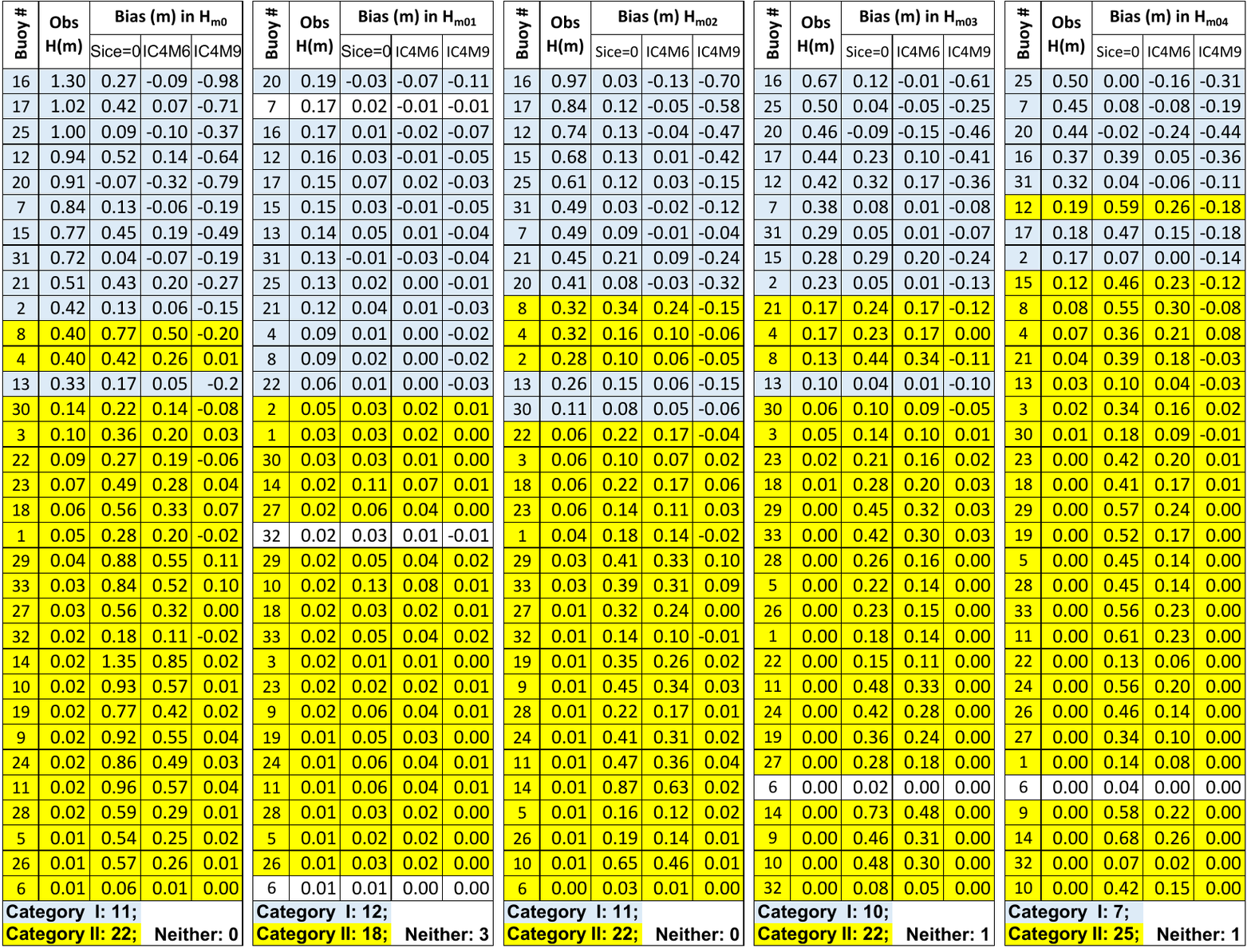}
\end{sidewaystable}

Buoy locations and times are provided to WW3, which would run and create a file tracking the buoys. The file contains WW3 2D spectra at buoy 
observation times at grid cells that bound the observation locations. A post-processing program is run to carry out the spatial interpolation and directional integration, thus producing the co-located model-buoy data points. 

The results from three model runs, using IC4M9,  IC4M6, and $S_{ice}=0$ (see B2, B3, B4 in Table \ref{table:cc-4AB}), are compared with the buoy observations. 
The model result with $S_{ice}=0$ primarily serves as an indicator of the ice effect since the biases in the modeled energies would be positive and high 
if the waves observed by a buoy were in sea ice. Statistical values of the model-observation comparison are 
computed for individual buoys, and the error metrics are tabulated in Table \ref{table:bias}. The biases (model $-$ observation) in $H_{m0}$ and 
$H_{m0n}$ for the four energy bands are selected for comparison. In each frequency band, the data are sorted in descending order of the observed 
wave energies. The data of ``Obs $H(\mbox{m})$ = 0.00" in each band have observed wave heights less than 3 mm.  
The data are also classified according to the relative magnitudes of the biases: Category I includes cases where IC4M6 gives smaller 
biases than IC4M9, while in Category II, IC4M9 outperforms IC4M6, having smaller biases. If the difference in the magnitudes of the biases for IC4M6 
and IC4M9 is less than 1 mm, the two dissipation functions are regarded to equally well estimate the observations, and the case does not belong 
to either Category I or II.  A few points can be made: 

(i) It is clearly evident that IC4M9 has a better skill than IC4M6 in estimating the SvalMIZ-24 data, across all frequencies.  
IC4M9 outperformed IC4M6 for 18, 22, 22, and 25 buoys in Bands 1, 2, 3 and 4, respectively, and for 22 buoys in estimating the total energy. 
On the other hand, the skill  of IC4M6 decreases with frequency, outperforming IC4M9 for 12, 11, 10 and 7 cases in Bands 1, 2, 3, and 4, 
respectively.  It is seen in Fig.~\ref{fig:ki-f} that at high frequencies the effect of $h_{ice}$ becomes increasingly significant 
according to IC4M9, while IC4M6 predicts insufficient dissipation, consistent with the results reported by Montiel et al. (2025).  
This corroborates the statistical evaluation (Table \ref{table:bias}) of IC4M9 and IC4M6 against the SvalMIZ-24 observations. 

(ii) For the cases in Category II, the buoy observed wave energies are very low, and mostly close to zero in Bands 3 and 4. Associated with those cases, 
the biases for $S_{ice}=0$, however,  are {\em positive} and large, indicating that wave damping by sea ice is strong in the observed 
data (but not modeled with $S_{ice}=0$). In contrast, the cases in Category I have high observed wave energies and are mostly associated 
with small biases for $S_{ice} =0$. This association between the observed energy and  biases for $S_{ice}=0$ suggests that 
the buoy measurements in Category II were mostly collected from waves that were strongly affected by sea ice, while the cases in Category I were 
for waves in areas with light ice coverage or of open water.  
From Fig.~\ref{fig:ki-f}, we expect IC4M9 to have a better skill than IC4M6 in predicting the strong dissipation in regions of heavy ice coverage and at  
high frequencies. This is confirmed by the comparison of biases. 

(iii) When IC4M9 underperformed in Category I, it gives negative biases for all the relevant buoys, and the magnitudes of the biases are especially large in  
Bands 2, 3, and 4. This means that IC4M9 strongly overestimates the dissipation in Category I. 
These systematic errors could imply that (a) there may be a calibration problem with IC4M9, or (b) the modeled ice forcing input to WW3 is overestimated. 
We did not re-calibrate IC4M9 since the SvalMIZ-24 dataset does not include $h_{ice}$ information along the buoy tracks, though $h_{ice}$ data at buoy 
deployments were collected (see Section \ref{sec:hice-disscusion}). On the other hand, it has been shown that IC4M9 can fit well other datasets without 
any re-calibration because it represents a scaling similitude (Yu et al., 2022). Therefore, we do not expect that the large {\em negative} biases 
for IC4M9 in Category I can be addressed by re-calibrating the formula; not to mention that a model lacks predictive skill if it needs to be re-calibrated 
in every application. The ice fields input to WW3 are ice concentration $a_{ice}$ and ice thickness $h_{ice}$. They are derived from the NAVGEM 0.18$^\circ$ 
product (Table \ref{table:cc-4AB}). Recall that the dissipation function is weighted by $a_{ice}$, i.e., the source term due to ice is $a_{ice}S_{ice}$. 
The model results using IC4M6 (which does not use  $h_{ice}$) do not indicate a significant overestimation of  $a_{ice}$: In Category II 
when IC4M6 underperformed, the biases are {\em positive} and large, meaning that IC4M6 predicts {\em insufficient} dissipation by sea ice, as expected. 
In Category I when IC4M6 performed better, the biases are overall small, with positive and negative values indicating non-systematic errors. 
It is therefore unlikely that the  modeled $a_{ice}$ is significantly over-predicted. This suggests  that the modeled  $h_{ice}$ provided to WW3 is likely biased high, 
penalizing IC4M9 and contributing to its under-estimation of wave energies in all bands in Category I. This is further discussed in Section \ref{sec:hice-disscusion}. 

\begin{figure}
\centering\includegraphics[width=4.75in]{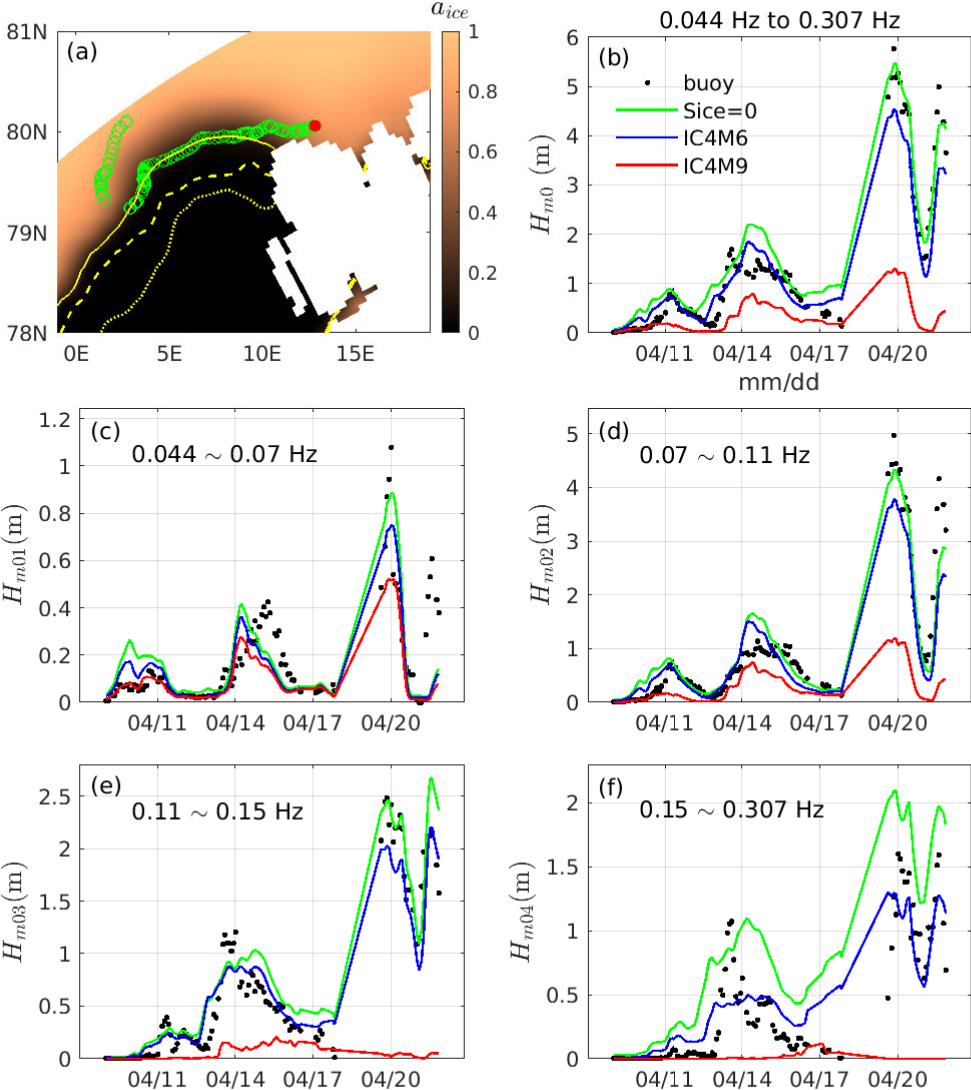}
\caption{Time series of wave measurements from buoy \#16. (a) Colormap: CICE $a_{ice}$ field at 21-Apr-2024 2100 UTC (end time). 
Yellow contours: $a_{ice}=0.15$ at 09-Apr-2024 0000 UTC (dotted), 17-Apr-2024 2100 UTC (dashed), and  21-Apr-2024 2100 UTC (solid). 
Green circles: buoy tracks. Red dot: start position at 5.27E 79.87N on 09 April 2024 at 00:30:00 UTC. 
(b)--(f) Time series of $H_{m0}$ for the total energy, and $H_{m0n}$ for the energy Band $n=1,2,3,4$. } 
\label{fig:buoy16}
\end{figure}

\begin{figure}
\centering\includegraphics[width=4.75in]{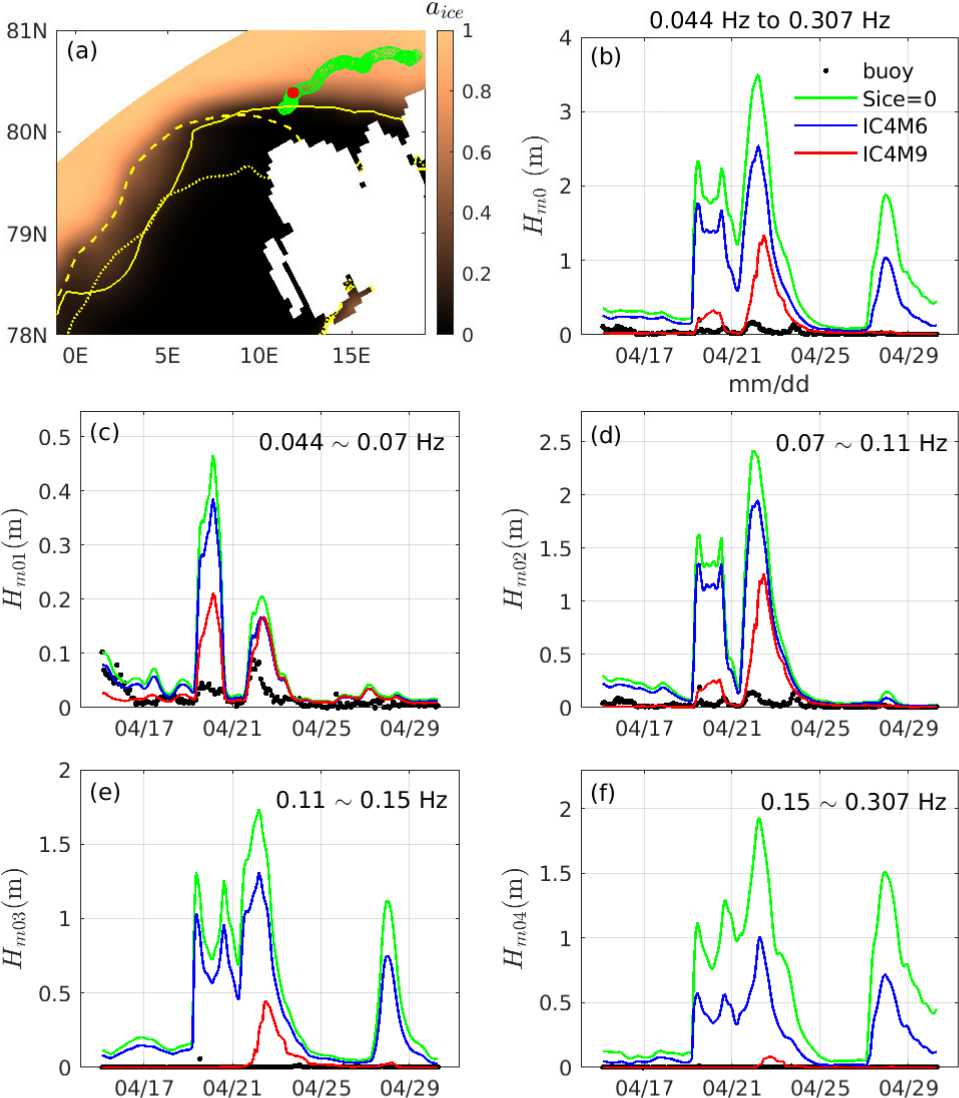}
\caption{Time series of wave measurements from buoy \#29.  (a) Colormap: CICE $a_{ice}$ field at 30-Apr-2024 0600 UTC (end time). 
Yellow contours: $a_{ice}=0.15$ at 15-Apr-2024 0300 UTC (dotted), 22-Apr-2024 2100 UTC (dashed), and 30-Apr-2024 0600 UTC (solid). 
Green circles: buoy tracks. Red dot: start position at 14.28E 80.65N on 15 April 2024 at 02:30:00 UTC.
(b)--(f) Time series of $H_{m0}$ for the total energy, and $H_{m0n}$ for the energy Band $n=1,2,3,4$.} 
\label{fig:buoy29}
\end{figure}

Examples of time series of $H_{m0}$ and $H_{m0n}$ are plotted in Figs.~\ref{fig:buoy16}--\ref{fig:buoy29} and compared to the observations. 
Buoy 16 (Fig.~\ref{fig:buoy16}) shows a typical case in Category I. 
The on-ice buoy drifted southwest during 09 to 18 April and turned north afterwards.  There are missing measurements from 17 April 2000 UTC to 
19 April 1500 UTC.  The ice edge was fast retreating polar-ward, as  seen from the receding contour of ice concentration $a_{ice} = 0.15$ 
and the widening area of open water; see Fig.~\ref{fig:buoy16}(a). There was a storm coming from the south with strong winds during the data-collection period. 
The temperature was down to $-25^\circ$C during the cold air outbreak, and barely 0$^\circ$C during the warm air intrusion event. Thus, excessive warming 
and fast ice melting were not likely. We expect that the polar-retreat of the ice edge was caused by the storm event, rather than ice melting so early in the spring.  
During the first two days of the deployment (up to 11 April), the buoy was in ice and sufficiently away from the open water,  recording very low wave energy 
in all frequency bands. The results given by IC4M9 for this early period agree the best with the measurements, while IC4M6 and $S_{ice}=0$ slightly 
overestimate the observations. 
As the buoy approached the fast receding ice edge, high energy peaks occurred around 11-12 April, 15-16 April and  19-20 April, indicating 
that the buoy was in areas of light ice coverage or open water. The model results by IC4M6 and $S_{ice}=0$ are similar, both more skillful in predicting those 
energy peaks than IC4M9 which mostly underestimates the observations after 11 April. The progressively deteriorating performance of IC4M9 at high frequencies 
again signals  the negative impact of modeled $h_{ice}$. 

Buoy 29 (Fig.~\ref{fig:buoy29}) is representative of cases in Category II, where IC4M9 strongly outperforms IC4M6 and the high {\em positive} biases given by 
$S_{ice}=0$ signify that substantial ice dissipation occurs. 
The buoy drifted northeast, farther into the sea ice. The measured wave energies are low, and practically zero in Bands 3 and 4, during the long record of over 15 days.  
While the observed total wave height is only of a few centimeters, $H_{m0}$ given by IC4M6 can be over 2 m, and even higher by $S_{ice}=0$. 
IC4M9 also overestimates the wave heights at some occasions (see the events on 19-20 April and 21-22 April in Fig.~\ref{fig:buoy29}), 
but overall it compares better with the observations than IC4M6 does. 

\section{On the modeled and observed ice thickness in the MIZ} 
\label{sec:hice-disscusion}

As discussed above, the over-estimation of wave energy dissipation by IC4M9 may be attributed to the over-predicted ice thickness that it uses. 
Here we attempt an evaluation of the modeled $h_{ice}$ from GOFS 3.1 database (Section \ref{sec:ww3-barents}). 
GOFS 3.1 uses the ice model CICE with assimilation of ice concentration observations (Metzger et al., 2017). 
The accuracy of CICE $h_{ice}$ is not well established (e.g., Xu and Li 2023). There are many routine satellite-derived ice concentration and ice edge products, 
but relatively few products for ice thickness, and those that do exist are relatively new (CryoSat-2 altimeter, SMOS passive microwave radiometer). 
We are not aware of any existing evaluations of the accuracy of CICE $h_{ice}$ within the MIZ, which is our primary concern here. 

During the SvalMIZ-24 campaign, in situ $h_{ice}$ data were collected at the buoy deployments but not along the buoy tracks. 
When each buoy was deployed on a floe, a hole was drilled in the floe and the snow and ice thickness were recorded.  
Thus, we have an in situ dataset of 33 buoy ice thicknesses for comparison.  We note that these measurements of $h_{ice}$ are highly local (one drill hole each) 
and may not be representative of a larger area, e.g., a 9 km x 9 km grid cell (CICE resolution) or 6 km x 6 km grid cell (WW3 resolution). 
The drill holes were made near the center of each floe which may be the thickest part of the floe. 
The subjective process of selecting a floe for buoy deployment may also introduce bias.

We also include the SMOS $h_{ice}$ for comparison. The SMOS $h_{ice}$ is a relatively new, first-generation product (Kaleschke et al., 2012; 
Huntemann et al., 2014), 
and in general not expected to have the same level of accuracy as other more mature satellite products such as the ice areal fraction.  
The SMOS data are seasonably available, with the last product from the early April. 
Since the product is intended for the “freeze-up period” (Huntemann et al., 2014), the SMOS $h_{ice}$ is most suitable for sheet ice. 
In areas where sea ice concentration is considerably lower than 100\%, the SMOS retrieval algorithm can lead 
to a systematic negative bias in the estimated sea ice thickness (Tian-Kunze et al., 2014; M\"{u}ller et al., 2023).
Nevertheless, we use $h_{ice}$ as is reported by SMOS. The SMOS $h_{ice}$ is on  an approximately 12 km irregular grid, with one analysis per day. 

\begin{figure}
\centering 
\includegraphics[width=2.25in]{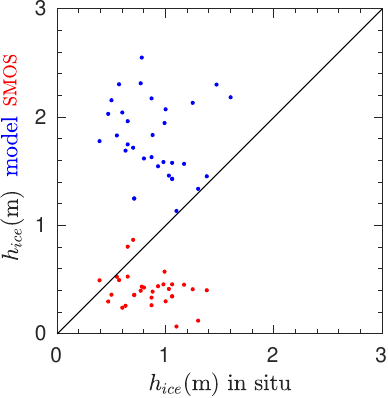}
\caption{Comparison of colocated $h_{ice}$ at buoy deployment: blue dots, $h_{ice,\,\mbox{model}}$ vs $h_{ics,\,\mbox{in situ}}$; 
red dots, $h_{ice,\,\mbox{\footnotesize SMOS}}$ vs $h_{ice,\,\mbox{in situ}}$.}
\label{fig:hice-33}
\end{figure} 

\begin{table} 
\caption{Statistical parameters for the comparisons of colocated $h_{ice}$ data at buoy deployment. 
Column 1 indicates the data pair to be compared, and $N$ is the number of colocations. 
The bias is computed for $a - b$. }\label{table:hice-33-statistics}
\begin{center}
\begin{tabular}{c| c | c c c }
\hline
  $a$ vs. $b$   & $N$ &Bias(m)   & RMSE(m)  & CC  \\
\hline         
 $h_{ice,\,\mbox{model}}$ vs. $h_{ice,\,\mbox{in situ}}$  & 33& 0.90  & 1.02  & -0.13  \\
 $h_{ice,\,\mbox{\footnotesize SMOS}}$ vs. $h_{ice,\,\mbox{in situ}}$ & 31&-0.44  & 0.55  & -0.27 \\ 
$h_{ice,\,\mbox{model}}$ vs. $h_{ice,\,\mbox{\footnotesize  SMOS}}$  & 31& 1.35  & 1.40 & 0.23 \\
\hline
\end{tabular}
\end{center}
\end{table}

The CICE $h_{ice}$ is first compared with in situ and SMOS observations for the 33 colocations (Fig.~\ref{fig:hice-33}). 
The SMOS product was missing on some days, so there are 31 colocated SMOS $h_{ice}$. 
At all but three of the colocations, we find that $h_{ice,\,\mbox{model}} > h_{ice,\,\mbox{in situ}} > h_{ice,\,\mbox{\footnotesize SMOS}}$, and for all colocations, 
$h_{ice,\,\mbox{model}} > h_{ice,\,\mbox{in situ}}$.  The statistical correlations among them are low, and the biases are high (Table \ref{table:hice-33-statistics}). 
While $h_{ice,\,\mbox{in situ}}$  is negatively correlated with both $h_{ice,\,\mbox{model}}$ and $h_{ice,\,\mbox{\footnotesize  SMOS}}$, 
there is a positive correlation between $h_{ice,\,\mbox{model}}$ and $h_{ice,\,\mbox{\footnotesize  SMOS}}$.  
The low correlations could be an indication that the in situ measurement of $h_{ice}$ may not be representative of the larger area around 
the buoy deployment, but the similarly low CC $=0.23$ between $h_{ice,\,\mbox{model}}$ and $h_{ice,\,\mbox{\footnotesize SMOS}}$ (both are representative of a larger area) 
does not support that argument.  Moreover, $h_{ice,\,\mbox{in situ}}$ is likely biased high, since thicker floes were selected for 
buoy deployments, and yet $h_{ice,\,\mbox{model}} >  h_{ice,\,\mbox{in situ}}$ at all locations. 
This suggests that the CICE $h_{ice}$ is almost certainly biased high against the unkown true ice thickness at those locations. 

\begin{figure}
\centering 
\includegraphics[width=4.75in]{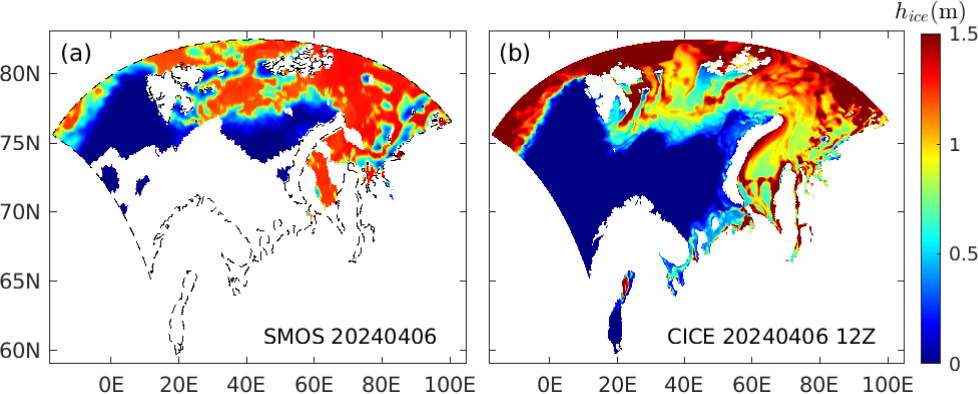}
\caption{(a) SMOS $h_{ice}$ remapped to the  6-km WW3 Barents Sea grid for 6 April 2024. 
The black dashed lines marks the land boundaries. The white color at the model sea points indicates that no valid information is available for that day. 
(b) CICE $h_{ice}$ from GOFS 3.1 remapped to the  WW3 grid for 1200 UTC 6 April 2024.}
\label{fig:hice-2D}
\end{figure}

\begin{figure}
\centering 
\includegraphics[width=4.75in]{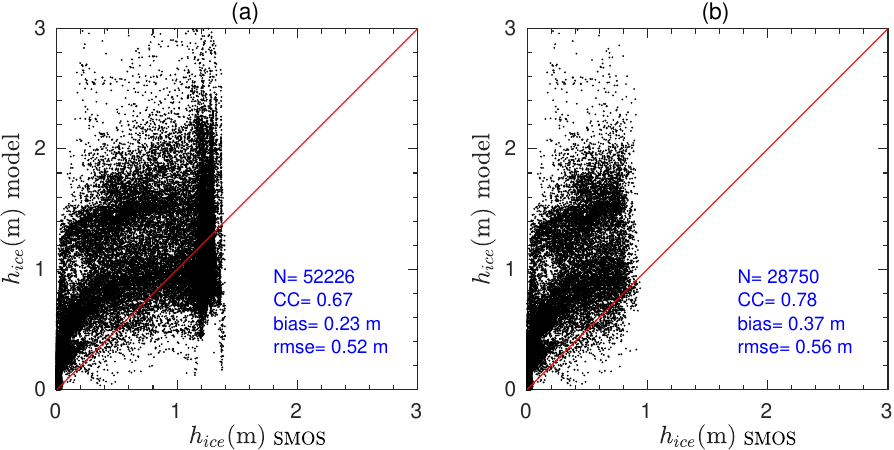}
\caption{Comparison of the remapped CICE and SMOS $h_{ice}$ for 6 April 2024. 
(a) Scatter plot of $N=52226$ colocations from the non-quality controlled data. 
(b) Scatter plot for $N=28750$ colocations from the  quality controlled data (i.e., saturation ratio $\leq$ 75\% ).   } 
\label{fig:hice-2D-scatter}
\end{figure}

We then examine the CICE and SMOS $h_{ice}$ over the Barents Sea model domain to have a broader spatial view of the  differences. 
Since the SMOS field is available daily, we compare the CICE and SMOS  $h_{ice}$ on the same day at 1200 UTC. The SMOS product and 
GOFS/CICE output are remapped (via spatial interpolations) to the Barents Sea WW3 6-km grid, and their colocations are identified.  
Examples of remapped SMOS and CICE $h_{ice}$ are shown in Fig.~\ref{fig:hice-2D}. Note the significantly higher CICE $h_{ice}$ in the 
west of Svalbard where most buoy tracks were during the SvalMIZ-24 campaign. 
Examples of scatter plots are given in Fig.~\ref{fig:hice-2D-scatter} for 06 April 2024 1200 UTC. There are 52226 colocations. 
With a quality control (QC; e.g., the max saturation ratio of  75\%), the colocations are reduced to 28750. 
The correlation coefficient is increased from 0.67 to 0.78 by applying the QC, justifying the application.  
These are significantly higher than CC $=0.23$ for the 33 colocations at buoy deployment. 
The apparent discrepancy is likely attributed to the fact that the colocations at buoy deployment are in the MIZ fairly close to the ice 
edge (see Figs.~\ref{fig:buoy16}--\ref{fig:buoy29}), whereas the comparison in Fig.~\ref{fig:hice-2D-scatter} includes many points farther into the ice pack. 
The biases of $h_{ice,\,\mbox{model}}-h_{ice,\,\mbox{\footnotesize SMOS}}$ is  0.37 m with the QC, while for the colocations at buoy deployments 
it is substantially higher (1.35 m). In any case, the CICE $h_{ice}$ field is much thicker than that from SMOS in the MIZ. An over-prediction of ice thickness 
would result in an over-prediction of ice dissipation by IC4M9, while not affecting IC4M6.  The penalty on IC4M9  is most severe near the ice edge where 
the fastest exponential decay of the wave amplitude occurs, and less pronounced farther away from the ice edge as the wave amplitude diminishes into the ice field. 
(Recall that the amount of wave energy dissipated is proportional to the local magnitude of the energy itself.) This is consistent with the finding in 
Section \ref{sec:obs-buoy} that IC4M9 systematically, strongly overestimates the dissipation in Category I which includes cases of waves in areas with light ice coverage. 

These comparisons have cautionary implications for estimation of dependence of $S_{ice}$ on $h_{ice}$. This was already remarked upon 
in Rogers et al. (2021b) and Yu et al. (2022) but is worth restating here. Hypothetically, if $h_{ice}$ from SMOS and CICE is thinner and thicker, respectively, relative to the unknown true ice thickness, use of either will bias any calibration of that dependence on $h_{ice}$.

\section{Summarizing remarks}
\label{sec:summary}

We have presented the recent advances in modeling ocean waves in the Arctic MIZ using the system developed for operational use by the U.S. Navy. 
We have focused our discussions on the applications in the extended Barents Sea, evaluating the performance 
skills of two formulations of wave attenuation by sea ice that are used by default in versions of the Navy's modeling system: the new formulation  
IC4M9 in equation (\ref{eq:ic4m9}) that describes wave attenuation depending on the ice thickness in association with the dependence on the wave frequency 
versus the previous default option IC4M6 which omits the dependence on ice thickness. 
The modeling results are validated against the observed wave spectra from the SWIM/CFOSAT product and the SvalMIZ-24 buoys.  
Our main conclusions are as follows. 

(i) When compared with the SWIM data, the model results using different ice dissipation schemes, including $S_{ice}=0$, show similar statistical 
performances, indicating that the colocated SWIM observations are mostly in regions of light sea ice coverage or open water.  
For such conditions, the high-latitude WW3 modeling has a good skill, with the model-observation correlations of 0.94 or higher for frequency bands 
above 0.08 Hz, and of 0.98 for all frequency bands combined, provided that the input wind has a sufficiently high resolution (e.g., the 0.18$^\circ$ NAVGEM product) 
and the ocean surface current is included. For frequencies below 0.08 Hz, the correlations are somewhat reduced, presumably due to observational errors, 
model forcing, and/or model errors, unrelated to sea ice effects.  Those causes are a topic for future investigations. 

(ii) Evaluating against the SvalMIZ-24 buoy wave observations, IC4M9 outperformed IC4M6 in all frequency bands and 
the total energy (Table \ref{table:bias}). IC4M6 performed well mostly for cases where the buoy tracks were close to the ice edge or in open 
water, similar to the condition of thin ice for which IC4M6 is derived. By contrast, IC4M9 is derived from a dataset in the Antarctic sea ice, 
yet it performed well in the Arctic MIZ without a re-calibration. As shown in Yu et al. (2022), without any re-calibration, 
formula (\ref{eq:ic4m9}) can fit well other datasets from various independent field and laboratory studies, suggesting that it represents a scaling similitude
because of the normalization applied in its derivation. The finding here offered yet another piece of evidence. 

(iii) When IC4M9 underperformed, it consistently under-predicts the observed wave energies, suggesting systematic errors. Our diagnostic analysis 
shows that the CICE ice thickness field provided to IC4M9 is  biased high in the MIZ, thus penalizing IC4M9 while not affecting IC4M6 which is 
frequency-dependent only. While this is disappointing, it reveals the importance of $h_{ice}$ for predicting waves in the MIZ. 
As an exploration, we have tested and found  that if we calibrate and reduce the CICE $h_{ice}$ using the observations, 
the performance of IC4M9 is improved when compared with the SvalMIZ-24 data (not shown here). 
If accurate and routine estimates of wave attenuation are available from satellite instruments and/or other types of observations, 
this sensitive response to $h_{ice}$ may be exploited in wave-ice coupled modeling to improve the prediction accuracy of $h_{ice}$ in the MIZ. 
To  facilitate successful future applications, we recommend broader testing of IC4M9. 

\section*{Acknowledgments} 
The CNES (Centre national d'études spatiales) provided the SWIM/CFOSAT dataset. The production of the SMOS sea ice thickness data was funded by the ESA project ``SMOS \& CryoSat-2 Sea Ice Data Product Processing and Dissemination Service", and data from April 6 to 15 2024 were obtained from \url{https://www.meereisportal.de} (grant: REKLIM-2013-04). 
A. Carrasco gratefully acknowledge funding from the MakingWaves project from the Norwegian Research Council (grant no. NFR-325654).

This is NRL contribution NRL/JA-7320-25-6982 and is approved for public release.

\section*{Availability Statement}

The OpenMetBuoy observations from the SvalMIZ-24 campaign are available on GitHub: 
\url{https://github.com/jerabaul29/2025_Svalbard_MIZ_KVS_SvalMIZ24}.
The model data used in this study are available at: \url{https://data.mendeley.com/datasets/9myzt338gv/1}.  

The newest WW3 source code including the IC4M9 model and related document are available on GitHub: 
\url{https://github.com/NOAA-EMC/WW3/tree/develop/model/src}, 
\url{https://github.com/NOAA-EMC/WW3/blob/develop/model/src/w3sic4md.F90}, 
\url{https://github.com/NOAA-EMC/WW3/blob/develop/manual/eqs/ICE4.tex}.



\begin{thebibliography}{100}
\bibitem[Ackley etal. (2020]{Ackley-etal-2020}
   {Ackley, S. F., Stammerjohn, S., Maksym, T., Smith, M., Cassano, J., Guest, P., et al.} (2020).
   {Sea-ice production and air/ice/ocean/ biogeochemistry interactions in the Ross Sea during the PIPERS 2017 autumn field campaign.}
   {\em  Annals of Glaciology}, {\bf 61}(82), 181--195.
\bibitem[Aouf etal. (2019)]{Aouf-etal-2019}
   {Aouf, L., Dalphinet, A., Hauser, D., Delaye, L., Tison, C., Chapron, B.,  et al.} (2019). 
   {On the assimilation of CFOSAT wave data in the wave model MFWAM: Verification phase.}
   {\em 2019 IEEE international geoscience and remote sensing symposium (IGARSS 2019)}, Yokohama, Japan. 2019, pp. 7959--7961. 
   https://doi.org/10.1109/IGARSS.2019.8900180
\bibitem[Aouf etal. (2021)]{Aouf-etal-2021}
  {Aouf, L., Hauser, D., Chapron, B., Toffoli, A., Tourain, C., \& Peureux, C.}  (2021). 
  {New directional wave satellite observations: Towards improved wave forecasts and climate description in Southern Ocean.} 
  {\em Geophysical  Research Letters}, {\bf 48}, e2020GL091187. https://doi.org/10.1029/2020GL091187.      
\bibitem[Ardhuin et al. (2018)]{Ardhuin-etal-2018}
  {Ardhuin, F., Boutin, G., Stopa, J., Girard-Ardhuin, F., Melsheimer, C.,  Thomson, J., Kohout, A.,  Doble, M., \& Wadhams, P.} (2018). 
  {Wave Attenuation Through an Arctic Marginal Ice Zone on 12 October 2015: 2. Numerical Modeling of Waves and Associated Ice Breakup}. 
  {\em Journal of Geophysical Research: Oceans}, {\bf 123}(8), 5652--5668.  
\bibitem[Bennetts et al. (2012)]{Bennetts-etal-2012} 
 {Bennetts, L. G., \&  Squire, V.  A.} (2012). {On the calculation of an attenuation coefficient for transects of ice-covered ocean}. 
 {\em Proceedings of the Royal Society A}, {\bf 468}, Issue 2137, 136--162. 
\bibitem[Booij et al. (1999)]{Booij-etal-1999}
  {Booij, N., Ris, R. C., \& Holthuijsen, L. H.} (1999). {A third-generation wave model for coastal regions, Part 1: Model description and validation.} 
  {\em Journal of Geophysical Research}, {\bf 104}(C4), 7649--7666.
\bibitem[Boutin etal. (2018)]{Boutin-etal-2018}  
  {Boutin, G.,  Ardhuin, F., Dumont, D.,  S\"{e}vigny, C., Girard-Ardhuin, F. \& Accensi, M.} (2018). 
  {Floe Size Effect on Wave-Ice Interactions: Possible Effects, Implementation in Wave Model, and Evaluation}. 
  {\em Journal of Geophysical Research: Oceans}, {\bf 123}, 4779--4805.    
\bibitem[Boutin etal. (2022)]{Boutin-etal-2022}
  {Boutin, G., Williams, T., Horvat, C., \& Brodeau, L.} (2022). {Modelling the Arctic wave-affected marginal ice zone: a comparison with ICESat-2 observations.}
  {\em  Philosophical Transactions of the Royal Society A}, {\bf  380}, 20210262. https://doi.org/10.1098/rsta.2021.0262    
\bibitem[Boutin etal. (2021)]{Boutin-etal-2021}
  {Boutin, G., Williams, T., Rampal, P., Olason, E., \& Lique, C.} (2021). {Wave--sea-ice interactions in a brittle rheological framework.} 
  {\em The Cryosphere}, {\bf 15}, 431--457.   
\bibitem[Campbell etal. (2010)]{Campbell-etal-210}
 {Campbell, T. J.,  Allard, R. A., Preller, R., Smedstad, L., Wellcraft, A., Chen, S., et al.} (2010). {Integrated Modeling of the Battlespace Environment.} 
 {\em  Computing in Science \& Engineering}, {\bf 12}, no. 5, 36--45.  doi:10.1109/MCSE.2010.78.      
\bibitem[Chassignet etal. (2003)]{Chassignet-etal-2003}
  {Chassignet, E. P., Smith, L. T., Halliwell, G. R., \& Bleck, R.} (2003). {North Atlantic simulations with the HYbrid Coordinate Ocean Model (HYCOM): 
  Impact of the vertical coordinate choice,  reference pressure, and thermobaricity}. 
  {\em Journal of Physical Oceanography}, {\bf 33}, 2504--2526. 
\bibitem[Chen etal. (2019)]{Chen-etal-2019}
   {Chen, H., Gilbert, R. P., \& Guyenne, P.} (2019). {Dispersion and attenuation in a porous viscoelastic model for gravity waves on an ice-covered ocean.} 
   {\em  European Journal of Mechanics - B Fluids},  {\bf 78}, 106--114. 
\bibitem[Collard et al. (2015)]{Collard-etal-2015}
   {Collard, F., Quartly, G., Konik, M., Johannessen, J., Korosov, A., Chapron, B., et al.} (2015). 
   {Ocean virtual Laboratory: A new way to explore multi-sensor synergy demonstrated over the Agulhas region.} 
   {\em Proceedings of Sentinel-3 for Science Workshop}, ESA Workshop, Venice, Italy, 1-5 June 2015. 
\bibitem[Collins and Rogers (2017)]{Collins-Rogers-2017} 
    {Collins, C. O., \&  Rogers, W. E.} (2017). {A source term for wave attenuation by sea ice in WAVEWATCH III: IC4}. 
    {\em NRL Memorandum Report}, NRL/MR/7320--17-9726, 25 pp.   https://www7320.nrlssc.navy.mil/pubs/2017/rogers-2017.pdf   
\bibitem[Doble and Bidlot (2013)]{Doble-Bidlot-2013}
  {Doble, M.J., Bidlot, J.-R.} (2013). 
  {Wavebuoy measurements at the Antarctic sea ice edge compared with an enhanced ECMWF WAM: progress towards global waves-in-ice modeling.} 
  {\em Ocean Modelling}, {\bf 70}, 166--173, doi: 10.1016/j.ocemod.2013.05.012.  
\bibitem[Doble et al. (2015)]{Doble-etal2015}
  {Doble, M. J., De Carolis, G., Meylan, M. H., Bidlot, J.-R., \& Wadhams, P.} (2015).  
  {Relating wave attenuation to pancake ice thickness, using field measurements and model results}.
  {\em Geophysical Research Letters}, {\bf 42}, 4473--4481.   
\bibitem[Doyle etal. (2014)]{Doyle-etal-2014}  
  {Doyle, J. D., Hodur, R. M.,  Chen, S.,  Jin, Y.,  Moskaitis, J. R.,  Wang, S., et al.} (2014). 
   {Tropical cyclone prediction using COAMPS-TC.}  
   {\em Oceanography},  {\bf 27}(3), 104--115, http://dx.doi.org/10.5670/oceanog.2014.72  
\bibitem[Dumas-Lefebvre etal. (2023)]{Dumas-etal-2023}
  {Dumas-Lefebvre, E., \& Dumont, D.} (2023). {Aerial observations of sea ice breakup by ship waves.} 
  {\em The Cryosphere}, {\bf 17}, 827--842.     
\bibitem[Dumont et al. (2011)]{Dumont-etal-2011}
  {Dumont, D., Kohout, A., \&  Bertino, L.} (2011). {A wave-based model for the marginal ice zone including a floe breaking parameterization}. 
  {\em Journal of Geophysical Research}, {\bf 116}, C4, C0400. 
\bibitem[Hauser etal. (2021)]{Hauser-etal-2021}
  {Hauser, D., Tourain, C., Hermozo, L., Alraddawi, D., Aouf, L., Chapron, B.,  et al.} (2021). 
  {New observations from the SWIM radar on-board CFOSAT: Instrument validation and ocean wave measurement assessment.} 
  {\em  IEEE Transactions on Geoscience and Remote Sensing}, {\bf 59}, NO 1, 5--26, https://doi.org/10.1109/TGRS.2020.2994372  
\bibitem[Hayhoe etal. (2018)]{Hayhoe-etal-2018}
  {Hayhoe, K., Wuebbles, D. J., Easterling, D. R.,  Fahey, D. W., Doherty, S., Kossin, J.,  et al.} (2018). 
  {Our Changing Climate. In: Reidmiller, D. R., Avery, C. W., Easterling, D. R., Kunkel, K. E.,  Lewis, K. L. M., Maycock, T. K., Steward, B. C. (Eds.)},   
   {\em Impacts, Risks, and Adaptation in the United States: Fourth National Climate Assessment, Volume II}.  
  U.S. Global Change Research Program, Washington, DC, USA, 63--135, doi: 10.7930/NCA4.2018.CH2
\bibitem[Hogan etal. (2014)]{Hogan-etal-2014}
   {Hogan, T., Liu, M., Ridout, J. A., Peng, M. S., Whitcomb, T. R., Ruston, B. C., et al.}  (2014). {The Navy Global Environmental Model.} 
   {\em Oceanography}, {\bf 27}(3), 116-125. 
\bibitem[Horvat and Tziperman (2015)]{Horvat-Tziperman-2015}
   {Horvat, C., \&  Tziperman, E.} (2015). {A prognostic model of sea-ice floe size and thickness distribution}. 
   {\em The Cryosphere}, {\bf 9}, 2119--2134.    
\bibitem[Horvat etal. (2020)]{Horvat-etal-2020}
   {Horvat, C., Blanchard-Wrigglesworth, E., \& Petty, A. A.} (2020). {Observing waves in sea ice with ICESat-2.}
   {\em Geophysical Research Letters}, {\bf 47}, e2020GL087629. https://doi.org/10.1029/2020GL087629     
\bibitem[Hunke and Lipscomb (2018)]{Hunke-Lipscomb-2018} 
  {Hunke, E. C.,  \& Lipscomb, W. H.} (2010). {CICE: The Los Alamos Sea Ice Model. Documentation and Software User's Manual. Version 4.1.} 
  {\em Technical Report}, LA-CC-06-012. T-3 Fluid Dynamics Group, Los Alamos National Laboratory. 
\bibitem[Huntemann etal. (2014)]{Huntemann-etal-2014}
  {Huntemann, M., Heygster, G., Kaleschke, L., Krumpen, T., Mäkynen, M., \& Drusch, M.} (2014).
  {Empirical sea ice thickness retrieval during the freeze-up period from SMOS high incident angle observations.}
  {\em The Cryosphere}, {\bf 8}, 439--451, doi:10.5194/tc-8-439-2014.  
\bibitem[Kaleschke etal. (2012)]{Kaleschke-etal-2012}
  {Kaleschke, L., Tian-Kunze, X.,  Maa\ss, N.,  M\"{a}kynen, M., \& Drusch, M.} (2012). 
  {Sea ice thickness retrieval from SMOS brightness temperatures during the Arctic freeze-up period.}
  {\em Geophysical Research Letters}, {\bf 39}, L05501, doi:10.1029/2012GL050916.    
\bibitem[Kaleschke etal. (2016)]{Kaleschke-etal-2016}
  {Kaleschke, L., Tian-Kunze, X., Maa\ss, N., Beitsch, A., Wernecke, A., Miernecki, M.,  et al.} (2016). 
  {SMOS sea ice product: Operational application and validation in the Barents Sea marginal ice zone.} 
  {\em Remote Sensing of Environment}, {\bf 180}, 264--273, https://doi.org/10.1016/j.rse.2016.03.009.
\bibitem[Keller (1998)]{Keller-1998}  
  {Keller, J. B.} (1998). 
  {Gravity waves on ice-covered water}. 
  {\em Journal of Geophysical Research: Oceans}, {\bf 103}, 7663--7669.
\bibitem[Kohout etal. (2016)]{Kohout-etal-2016}
  {Kohout, A. L., Williams, M. J. M., Toyota, T., Lieser, J., \& Hutchings, J.} (2016). {In situ observations of wave-induced sea ice breakup.} 
  {\em Deep-Sea Research II}, {\bf 131}, 22--27.  
\bibitem[LeMerle etal. (2021)]{LeMerle-etal-2021}  
  {Le Merle, E., Hauser, D., Peureux, C., Aouf, L., Schippers, P., Dufour, C., \& Dalphinet, A.} (2021). 
  {Directional and frequency spread of surface ocean waves from SWIM measurements.}
  {\em Journal of  Geophysical Research: Oceans}, {\bf 126}, e2021JC017220. https://doi.org/10.1029/2021JC017220     
\bibitem[Liu et al. (2020)]{Liu-etal-2020}
 {Liu, Q., Rogers, W. E., Babanin, A., Li, J., \& Guan, C.} (2020). {Spectral Modeling of Ice-Induced Wave Decay}.
 {\em Journal of Physical Oceanography}, {\bf 50}(6), 1583--1604.   
\bibitem[Metzger etal. (2017)]{Metzger-etal-2017}
   {Metzger, E. J., Helber, R. W., Hogan, P. J., Posey, P. G., Thoppil, P. G., Townsend, T. L., et al.} (2017). 
   {Global Ocean Forecast System 3.1 validation testing.}  
   {\em Naval Research Laboratory Techical  Report}.  NRL/MR/7320–17-9722, 56 pages. https://apps.dtic.mil/sti/citations/AD1034517.   
\bibitem[Meylan etal. (2014)]{Meylan-etal-2014}
   {Meylan, M. H., Bennetts, L. G., \& Kohout, A. L.} (2014). {In situ measurements and analysis of ocean waves in the Antarctic marginal ice zone.} 
   {\em Geophysical Research Letters}, {\bf 41}, 5046--5051.
\bibitem[Meylan etal. (2021)]{Meylan-etal-2021}
   {Meylan, M. H., Horvat, C., Bitz, C. M., \& Bennetts, L. G.} (2021). {A floe size dependent scattering model in two- and three-dimensions for wave 
   attenuation by ice floes.} 
   {\em Ocean Modelling}, {\bf 161}, 101779.    
\bibitem[Mokus and Montiel (2022)]{Mokus-Montiel-2022}
  {Mokus, N. G. A., \&  Montiel, F.} (2022). {Wave-triggered breakup in the marginal ice zone generates lognormal floe size distributions: a simulation study.}   
  {\em The Cryosphere}, {\bf 16}, 4447--4472. https://doi.org/10.5194/tc-16-4447-2022   
\bibitem[Montiel et al. (2025)]{Motiel-etal-2025}
   {Montiel, F., Forbes, M., Echevarria, M., Rapizo, H., \& Gamble, C.} 2025. 
   {Evaluation of ice dissipation parameterizations in spectral ocean wave model WAVEWATCH III: An intercomparison analysis.} 
   {\em Journal of Geophysical Research: Oceans}, {\bf 130}, e2024JC022113. https://doi.org/10.1029/2024JC022113    
\bibitem[Muller etal. (2023)]{Muller-etal-2023}
  {M\"{u}ller, M., Batrak, Y., Dinessen, F., Grote, R., \& Wang, K.} (2023).
  {Challenges in the description of sea ice for a kilometer-scale weather forecasting system.}
  {\em Weather and Forecasting}, {\bf  38}, 1157--1171. https://doi.org/10.1175/WAF-D-22-0134.1  
\bibitem[Muller etal. (2024)]{Muller-etal-2024}
 {M\"{u}ller, M., Rabault, J.,  \& Palerme, C.}  (2024).  {Svalbard Marginal Ice Zone 2024 campaign--Cruise report.} 
 arXiv, 2407.18936, https://doi.org/10.48550/arXiv.2407.18936.   
\bibitem[Muller etal. (2025)]{Muller-etal-2025}
   {M\"{u}ller, M., Rabault, J., Abdel-Fattah, D.,  \& Sutherland, G.} (2025). 
   {Distributed Observation Networks in the Arctic Marginal Ice Zone to Advance Forecasting Systems.} 
   {\em  Bulletin of the American Meteorological Society}, {\bf 106}(6), E1204-E1210. https://doi.org/10.1175/BAMS-D-25-0082.1  
\bibitem[Nose etal. (2024)]{Nose-etal-2024}
   {Nose, T.,  Katsuno, T., Waseda, T.,  Ushio, S., Rabault, J., Kodaira, T., et al.}  (2024). 
   {Observation of wave propagation over 1,000 km into Antarctica winter pack ice.}
   {Coastal Engineering}, {\bf 66}(1), 115--131.  doi: 10.1080/21664250.2023.2283243   
\bibitem[Prinsenberg and Peterson (2011)]{Prinsenberg-Peterson-2011}
    {Prinsenberg, S. J., \& Peterson, I. K.} (2011). 
    {Observing regional-scale pack-ice decay processes with helicopter-borne sensors and moored upward-looking sonars.}
    {\em Annals of Glaciology}, {\bf 52}(57), 35--42.   
\bibitem[Rabault etal. (2022)]{Rabault-etal-2022}   
  {Rabault, J., Nose, T., Hope, G., M\"{u}ller, M., Breivik, {\O}., Voermans, J., et al.}  (2022). 
  {OpenMetBuoy-v2021: An Easy-to-Build, Affordable, Customizable, Open-Source Instrument for Oceanographic Measurements of Drift and Waves
   in Sea Ice and the Open Ocean.} 
   {\em Geosciences}, {\bf 12}(3), 110, https://doi.org/10.3390/geosciences12030110         
\bibitem[Rabault etal. (2024)]{Rabault-etal-2024}
   {Rabault, J., Halsne, T., Carrasco, A., Korosov, A., Voermans, J., Bohlinger, P., et al.} (2024)
   {Buoy measurements of strong waves in ice amplitude modulation: a signature of complex physics governing waves in ice attenuation}. 
   {arXiv:2401.07619v2}. https://doi.org/10.48550/arXiv.2401.07619
\bibitem[Roach etal. (2018)]{Roach-etal-2018}
   {Roach, L. A., Horvat, C., Dean, S. M., \& Bitz, C. M.} (2018). {An emergent sea ice floe size distribution in a global coupled ocean-sea ice model.}
   {\em Journal of Geophysical Research: Oceans}, {\bf 123}, 4322--4337.  https://doi.org/10.1029/2017JC013692      
\bibitem[Rogers and Orzech (2013)]{Rogers-ORzech-2013}
   {Rogers, W. E.,  \& Orzech, M.}  (2013). {Implementation and testing of ice and mud source functions in WAVEWATCH III\textsuperscript{\textregistered}.}
   {\em NRL Memorandum Report}, NRL/MR/7320-13-9462, 31 pp. Naval Research laboratory.  https://apps.dtic.mil/sti/tr/pdf/ADA584701.pdf
\bibitem[Rogers and Zieger (2014)]{Rogers-Zieger-2014}
   {Rogers, W.E., \& Zieger, S.}  (2014). {New wave-ice interaction physics in WAVEWATCH III.} 
   In {\em proceedings of 22nd IAHR International Symposium on Ice}, Singapore, 11-15 August 2014, 8 pages. 
\bibitem[Rogers et al. (2016)]{Rogers-etal-2016}
  {Rogers, W.E., Thomson, J., Shen, H. H., Doble, M. J.,  Wadhams, P., \& Cheng, S.} (2016). 
  {Dissipation of wind waves by pancake and frazil ice in the autumn Beaufort Sea.} 
  {\em Journal of  Geophysical Research: Oceans}, {\bf121}, 7991--8007, doi:10.1002/2016JC012251     
\bibitem[Rogers et al. (2018a)]{Rogers-etal-2018a}
   {Rogers, W. E., Posey, P.,  Li, L.,  \& Allard, R. A.}  (2018a). 
   {Forecasting and hindcasting waves in and near the marginal ice zone: Wave modeling and the ONR “Sea State” field experiment.} 
   {\em  NRL Report}, NRL/MR/7320--18-9786, 179 pp.  https://apps.dtic.mil/sti/pdfs/AD1050539.pdf 
\bibitem[Rogers et al. (2018b)]{Rogers-etal-2018b}
   {Rogers, W. E., Meylan, M. H.,  \& Kohout, A. L.}  (2018b). 
   {Frequency distribution of dissipation of energy of ocean waves by sea ice using data from Wave Array 3 of the ONR “Sea State” field experiment.}
   {\em  NRL Report},  NRL/MR/7322--18-9801, 25 pp. https://apps.dtic.mil/sti/trecms/pdf/AD1058347.pdf   
\bibitem[Rogers et al. (2021a)]{Rogers-etal-2-21a}
   {Rogers, W. E., Meylan, M. H., \& Kohout, A. L.}  (2021a). {Estimates of spectral wave attenuation in Antarctic sea ice, using model/data inversion.} 
   {\em Cold Regions Science and Technology}, {\bf 182}, 103198,  https://doi.org/10.1016/j.coldregions.2020.103198
\bibitem[Rogers etal. (2021b)]{Rogers-etal-2021b}
   {Rogers, W. E., Yu, J., \& Wang, D. W.} (2021b). {Incorporating dependencies on ice thickness in empirical parameterizations of wave dissipation by sea ice.} 
   {\em NRL Technical Report}, NRL/OT/7320-21-5145, 35 pp. https://arxiv.org/abs/2104.01246.    
\bibitem[Rogers etal. (2025)]{Rogers-etal-2025}
   {Rogers, W. E., Campbell, T. J., Yu, J., \& Allard, R. A.} (2025). {Uncoupled high-latitude wave model in COAMPS.} 
   {\em NRL Technical Report}, NRL/7320/MR--2025/1, 79 pages. https://doi.org/10.48550/arXiv.2508.16755 .        
\bibitem[Shen (2019)]{Shen-2019}
   {Shen, H. H.} (2019). {Modelling ocean waves in ice-covered seas}. {\em Applied Ocean Research}, {\bf 83}, 30--36.  
\bibitem[Smith etal. (2013)]{Smith-etal-2013}
   {Smith, T. A., Chen, S., Campbell, T.,  Martin, P.,  Rogers, W. E., Gabersek, S., et al.} (2013).
   {Ocean-wave coupled modeling in COAMPS-TC: A study of Hurricane Ivan (2004).} 
   {\em Ocean Modelling}, {\bf 69}, 181--194.   
\bibitem[Squire etal. (1995)]{Squire-etal-1995}
   {Squire, V. A., Dugan, J. P., Wadhams, P., Rottier, P. J., \& Liu, A. K.} (1995). {Of ocean waves and sea ice.}
   {\em  Annual Review of Fluid Mechanics},  {\bf 27},115--168.  
\bibitem[Squire (2018)]{Squire-2018}
   {Squire, V. A.} (2018). {A fresh look at how ocean waves and sea ice interact.}
   {\em  Philosophical Transactions of the Royal Society A },  {\bf 376}, 20170342. 
\bibitem[Stopa et al. (2016)]{Stopa-etal-2016}
   {Stopa, J. E., Ardhuin, F., \& Girard-Ardhuin, F.} (2016). {Wave climate in the Arctic 1992–2014: seasonality and trends}. 
   {\em  The Cryosphere}, {\bf 10}, 1605--1629.    
\bibitem[Stopa et al. (2018)]{Stopa-etal-2018}
   {Stopa, J. E., Sutherland, P., \& Ardhuin, F.} (2018). {Strong and highly variable push of ocean waves on Southern Ocean sea ice.}
   {\em  Proceedings of the National Academy of Sciences (PNAS)}, {\bf 115}, 5861--5865.        
\bibitem[SWIM (2019)]{swim-2019}
   {SWIM Production Users Guide.} (2019).  Centre National d'EtudeS Spatiales. Reference: CF-GSFR-MU-2530-CNES. 
\bibitem[Thomson (2022)]{Thomson-2022}
   {Thomson, J.} (2022). {Wave propagation in the marginal ice zone: connections and feedback mechanisms within the air--ice--ocean system.} 
   {\em  Philosophical Transactions of the Royal Society A }, {\bf 380}, 20210251.
\bibitem[Thomson and Rogers (2014)]{Thomson-Rogers-2014}
   {Thomson, J., \& Rogers, W. E.} (2014). {Swell and sea in the emerging Arctic Ocean.} 
   {\em Geophysical Research Letters}, {\bf 41}, 3136--3140, https://doi.org/10.1002/2014GL059983   
\bibitem[Thomson etal. (2017)]{Thomson-etal-2017}
   {Thomson, J., Ackley S., Shen, H. H., Rogers, W. E.} (2017). {The Balance of Ice, Waves, and Winds in the Arctic Autumn.} 
   {\em Eos}, {\bf 98}, 30--34. https://doi.org/10.1029/2017EO066029.
\bibitem[Tian-Kunze et al. (2014]{Tian-Kunze-etal-2014}
   {Tian-Kunze, X., Kaleschke, L., Maa\ss, N., M\"{a}kynen, M., Serra, N., Drusch, M.,  et al.} (2014).
   {SMOS-derived thin sea ice thickness: algorithm baseline, product specifications and initial verification.}
   {\em The Cryosphere}, {\bf 8}, 997--1018.  doi:10.5194/tc-8-997-2014   
\bibitem[Tolman (1991)]{Tolman-1991}
   {Tolman, H. L.} (1991). {A Third-generation model for wind-waves on slowly varying, unsteady, and inhomogeneous depths and currents.} 
   {\em Journal of Physical Oceanography},  {\bf 21}(6), 782--797.   
\bibitem[Voermans et al. (2020)]{Voermans-etal-2020}
   {Voermans, J. J., Rabault, J., Filchuck, K., Ryzhov, I., Heil, P., Marchenko, A., et al.} (2020). 
   {Experimental evidence for a universal threshold characterizing wave-induced sea ice break-up.}
   {\em The Cryosphere}, {\bf 14}, 4265--4278.  https://doi.org/10.5194/tc-14-4265-2020
\bibitem[Wadhams et al. (2018)]{wadhams-etal-2018}
   {Wadhams, P., Aulicino, G., Parmiggiani, F., Persson, P.O.G., \& Holt, B.} (2018). 
   {Pancake ice thickness mapping in the Beaufort Sea from wave dispersion observed in SAR imagery.} 
   {Journal of Geophysical Research: Oceans}, {\bf 123}, 2213--2237. https://doi.org/10.1002/2017JC013003 
\bibitem[Wang and Shen (2010)]{Wang-Shen-2010}
  {Wang, R., \& Shen, H. H.}  (2010). {Gravity waves propagating into an ice-covered ocean: A viscoelastic model.} 
  {\em Journal of  Geophysical Research}, {\bf 115}, C06024, doi:10.1029/2009JC005591.   
\bibitem[Williams et al. (2013)]{Williams-etal-2013}
  {Williams, T. D., Bennetts, L. D.,  Squire, V. A., Dumont, D., \& Bertino, L.} (2013). 
  {Wave-ice interactions in the marginal ice zone. Part 1: Theoretical foundations}. 
  {\em Ocean Modelling}, {\bf 70}, 81--91. 
\bibitem[WW3DG (2019)]{WW3DG-2019}  
   {The WAVEWATCH III\textsuperscript{\textregistered} Development Group (WW3DG).}  (2019). 
   {User manual and system documentation of WAVEWATCH III\textsuperscript{\textregistered} version 6.07.} 
   {\em Technical Note 333}, NOAA/NWS/NCEP/MMAB, College Park, MD, U.S.A., 465 pages + Appendices.      
\bibitem[Xu and Li (2023)]{Xu-Li-2023}
   {Xu, M., \& Li,  J.}  (2023). {Assessment of sea ice thickness simulations in the CMIP6 with CICE components.} 
   {\em  Frontiers in Marine Science}, {\bf  10}, 1223772. doi: 10.3389/fmars.2023.1223772         
\bibitem[Yang et al. (2024)]{Yang-et-al-2024}
   {Yang, C.-Y., Liu, J., \& Chen, D.} (2024). {Understanding the influence of ocean waves on Arctic sea ice simulation: a modeling study with an 
   atmosphere--ocean--wave--sea ice coupled model.} 
   {\em The Cryosphere}, {\bf 18}, 1215--1239.  
\bibitem[Yu et al. (2019)]{Yu-etal-2019}
   {Yu, J., Rogers, W. E., \& Wang, D.W.} (2019). {A scaling for wave dispersion relationships in ice-covered waters.} 
   {\em Journal of Geophysical Research: Oceans}, {\bf 124},  doi:10.1029/2018JC014870.
\bibitem[Yu et al. (2022)]{Yu-etal-2022}
   {Yu, J., Rogers, W. E.,  \& Wang, D.W.} (2022). {A new method for parameterization of wave dissipation by sea ice}. 
   {\em Cold Regions Science and Technology}, {\bf 199}, 103582, https://doi.org/10.1016/j.coldregions.2022.103582.  
\bibitem[Yu (2022)]{Yu-2022}
   {Yu, J.} (2022). {Wave boundary layer at the ice-water interface.} 
   {\em Journal of Marine. Science and Engineering}, {\bf 10}, 1472,   https://doi.org/10.3390/jmse10101472
\end{thebibliography}
\end{document}